\documentclass[a4paper,english,pagebackref]{article}
\usepackage{xspace}
\usepackage{amsthm,amsmath}
\usepackage{a4wide}
\usepackage{paralist}
\usepackage{booktabs}
\usepackage{authblk}
\usepackage{tabularx}
\usepackage[numbers,sort]{natbib}
\renewcommand{\citet}{\cite}
\renewcommand{\citeauthor}{\cite}

\makeatletter
\def\NAT@spacechar{~}%
\makeatother
\usepackage{tikz}
\usetikzlibrary{decorations.pathmorphing}
\usetikzlibrary{decorations.pathreplacing}
\usetikzlibrary{shapes,automata}

\usepackage{color}
\usepackage{graphicx}
\definecolor{mygrey}{rgb}{0.9,0.9,0.9}
\definecolor{darkgreen}{RGB}{0,100,0}

\usepackage{lipsum}
\usepackage{amsfonts}
\usepackage{graphicx}
\usepackage{epstopdf}
\usepackage{algorithmic}

\usepackage[T1]{fontenc}

\usepackage[pdfdisplaydoctitle,menucolor=orange!40!black,filecolor=magenta!40!black,urlcolor=blue!40!black,linkcolor=red!40!black,citecolor=green!40!black,colorlinks]{hyperref}
\hypersetup{pdftitle={Using Fractal-like Strutures in Cross-Compositions}, pdfauthor={Till Fluschnik, Danny Hermelin, Andr ́e Nichterlein, and Rolf Niedermeier}}
\usepackage[sort&compress,nameinlink,noabbrev,capitalize]{cleveref}

\theoremstyle{plain}
\newtheorem{theorem}{Theorem}
\newtheorem{corollary}{Corollary}
\newtheorem{definition}{Definition}
\newtheorem{proposition}{Proposition}

\newtheorem{observation}{Observation}
\newtheorem{remark}{Remark}

\newtheorem{lemma}{Lemma}
\newtheorem{constr}{Construction}

\crefname{constr}{Construction}{Constructions}
\crefname{step}{Step}{Steps}
\crefname{lemma}{Lemma}{Lemmas}
\crefname{observation}{Observation}{Observations}
\crefname{proposition}{Proposition}{Propositions}
\crefname{remark}{Remark}{Remarks}
\crefname{prop}{Property}{Properties}
\crefname{constr}{Construction}{Constructions}
\crefname{observation}{Observation}{Observations}
\crefname{remark}{Remark}{Remarks}
\crefname{theorem}{Theorem}{Theorems}
\Crefname{theorem}{Thm.}{Thms.}

\newcommand{\no}{\textsc{no}\xspace}
\newcommand{\yes}{\textsc{yes}\xspace}

\newcommand{\N}{\mathbb{N}}
\newcommand{\I}{\mathcal{I}}
\newcommand{\R}{\mathcal{R}}

\newcommand{\fractal}[1][T]{#1-fractal\xspace}
\newcommand{\Fractal}[1][T]{#1-Fractal\xspace}

\newcommand{\poly}{\ensuremath{\operatorname{poly}}}

\newcommand{\NP}{\ensuremath{\operatorname{NP}}}
\newcommand{\coNP}{\ensuremath{\operatorname{coNP}}}

\newcommand{\tw}{\ensuremath{\omega}}

\DeclareMathOperator{\dist}{dist}
\DeclareMathOperator{\diam}{diam}

\newcommand{\decprob}[3]{
	\begin{center}
		\begin{minipage}{0.95\textwidth}
			\noindent
			\textsc{#1}\\
			\setlength{\tabcolsep}{3pt}
			\begin{tabularx}{\textwidth}{@{}lX@{}}
					\normalsize \textbf{Input:} 		& \normalsize #2 \\
					\normalsize \textbf{Question:} 	& \normalsize #3
				\end{tabularx}
		\end{minipage}
	\end{center}
}

\newcommand{\spmve}{Length-Bound\-ed Edge-Cut\xspace}
\newcommand{\spmveAcr}{\textsc{LBEC}\xspace}
\newcommand{\spmveTsc}{\textsc{\spmve}\xspace}

\newcommand{\dmve}{Mi\-ni\-mum Di\-am\-e\-ter Edge Dele\-tion\xspace}
\newcommand{\dmveAcr}{\textsc{MDED}\xspace}
\newcommand{\dmveTsc}{\textsc{\dmve}\xspace}

\newcommand{\gmve}{Di\-rect\-ed Small Cy\-cle Transver\-sal\xspace}
\newcommand{\gmveAcr}{\textsc{DSCT}\xspace}
\newcommand{\gmveTsc}{\textsc{\gmve}\xspace}

\newcommand{\lv}{\sigma}
\newcommand{\rv}{\tau}

\newcommand{\thetitle}{Fractals for Kernelization Lower Bounds}

\title{
  {\thetitle{}}%
  \footnote{An extended abstract appeared in \textit{Proc.\ of the 43rd International Colloquium on Automata, Languages, and Programming (ICALP 2016)}.}
  }

\author[1]{Till~Fluschnik\thanks{Till~Fluschnik acknowledges support by the DFG, project DAMM (NI~369/13-2).}$ ^,$}
\author[2]{Danny~Hermelin\thanks{Danny~Hermelin was supported by a DFG Mercator fellowship within the project DAMM (NI~369/13-2) while staying at TU Berlin (August~2015). He has also %
received funding from the People Programme (Marie Curie Actions) of the European Union's Seventh Framework Programme (FP7/2007-2013) under REA grant agreement number 631163.11, and by the ISRAEL SCIENCE FOUNDATION (grant No. 551145/14).}$ ^,$}
\author[1]{Andr\'e~Nichterlein}
\author[1]{{Rolf~Niedermeier}}

\affil[1]{\small{Institut f\"ur Softwaretechnik und Theoretische Informatik, TU~Berlin, Germany, \texttt{\{till.fluschnik, andre.nichterlein, rolf.niedermeier\}@tu-berlin.de}}}
\affil[2]{\small{Department of Industrial Engineering and Management, Ben-Gurion University of the Negev, Israel, \texttt{hermelin@bgu.ac.il}}}

\date{}

\begin{document}

\maketitle

\begin{abstract}
  \noindent The composition technique is a popular method for excluding polynomial-size problem kernels for NP-hard parameterized problems. 
  We present a new technique exploiting triangle-based fractal structures for extending the range of applicability of com\-po\-si\-tions. 
  Our technique makes it possible to prove new no-polynomial-kernel results for a number of problems dealing with length-bounded cuts.
  In particular, answering an open question of Golovach and Thilikos [Discrete Optim.~2011], we show that, unless $\NP\subseteq \coNP/\poly$, the NP-hard \spmveTsc{}~(\spmveAcr) problem (delete at most~$k$ edges such that the resulting graph has no $s$-$t$ path of length shorter than~$\ell$) parameterized by the combination of $k$ and~$\ell$ has no polynomial-size problem kernel.
  Our framework applies to planar as well as directed variants of the basic problems and also applies to both edge and vertex deletion problems.
  Along the way, we show that \spmveAcr{} remains NP-hard on planar graphs, a result which we believe is interesting in its own right.
  
\end{abstract}

\section{Introduction}\label{sec:intro}
Lower bounds are of central concern all over computational complexity analysis.
With respect to fixed-parameter tractable problems~\cite{DowneyF13,FlumG06,Nie06,CyganFKLMPPS15}, currently there are two main streams in this context:%
\begin{enumerate}[(i)]%
	\item ETH-based lower bounds for the running times of exact algorithms~\cite{LMS11}
	  and
	\item lower bounds on problem kernel sizes; more specifically, the exclusion of poly\-nomial-size problem kernels~\cite{Kratsch14}.
\end{enumerate}
Both research directions for lower bounds rely on plausible complexity-theoretic assumptions, namely the Exponential-Time Hypothesis (ETH) and  $\NP\not\subseteq \coNP/\poly$, respectively.
In this work, we contribute to the second research direction, developing a new technique that exploits a triangle-based fractal structure in order to  exclude polynomial-size problem kernels (polynomial kernels for short) for edge and vertex deletion problems in the context of length-bounded cuts.

Kernelization is a key method for designing fixed-parameter algorithms~\cite{GN07,Kratsch14}; among all techniques of parameterized algorithm design, it has the presumably greatest potential for delivering practically relevant algorithms.
Hence, it is of key interest to explore its power and its limitations.
In a nutshell, the fundamental idea of kernelization is as follows.
Given a parameterized problem instance~$I$ with parameter~$k$, in polynomial time preprocess~$I$ by applying data reduction rules in order to simplify it and reduce it to an ``equivalent'' instance (the so-called (problem) kernel) of the same problem. 
For NP-hard problems, the best one can hope for is a problem kernel of size polynomial (or linear) in the parameter~$k$.
In a way, one may interpret kernelization (requested to run in polynomial time) as an ``exact counterpart'' of polynomial-time approximation algorithms. 
Indeed, linear-size problem kernels often imply constant-factor approximation algorithms~\cite[page 15]{Mar08}.
Approximation algorithmics has a highly developed theory (having produced concepts such as MaxSNP-hardness and the famous PCP~theory) for proving (relative to some plausible complexity-theoretic assumption) 
lower bounds on the approximation 
	factors~\cite{Vaz01,WS11,ACGKMP99}. 
We remark that there exist frameworks combining kernelization and approximation algorithms, namely $\alpha$-fidelity kernelization~\cite{FellowsKRS12} and lossy kernelization~\cite{LokshtanovPRS17}.

It is fair to say that in the younger field of kernelization the arsenal for proving lower bounds (particularly excluding polynomial kernels) so far is of smaller scope and needs further development.
The first results in this context were rather limited, and used approximation lower bounds~\cite{ChenFKX07}.
Following these, Bodlaender et al.~\citet{BodlaenderDFH09} (using a lemma by Fortnow and Santhanam~\citet{FortnowS11}) excluded polynomial kernels for several problems, such as \textsc{Longest Path} parameterized by solution size, under the assumption $\NP\not\subseteq \coNP/\poly$.
The core tool for showing these are so-called ``OR-compositions''.
The applicability of OR-compositions has meanwhile been developed further in several other works, e.g.~\cite{BodlaenderJK14,BTY11,DomLS14}.
Dell and van Melkebeek~\cite{DellM14} introduced a related framework that also allows to give lower bounds on the degree of the polynomial for problems admitting a polynomial kernel.
Finally, Drucker~\cite{Drucker15} recently showed that ``AND-compositions'' can also be used to exclude polynomial kernels.
Next, we discuss in some more detail OR-compositions. 
Roughly speaking, the idea behind an OR-composition for a parameterized problem is to encode the logical ``or'' of $t$~instances with parameter value~$k$ into a single instance of the same problem with parameter value~$k' = k^{O(1)} \log t$. %
In particular, given $t$~instances, the obtained instance is a yes-instance if and only if at least one of the given instances is a yes-instance. 
In this way an OR-composition can be viewed as a polynomial-time computable OR-gate.
An OR-composition for an NP-hard problem, along with a polynomial kernel for the same problem, implies $\NP\subseteq \coNP/\poly$~\cite{BodlaenderDFH09,FortnowS11}.

While for some problems, for example \textsc{Longest Path} with parameter solution size~\cite{BodlaenderDFH09}, a simple disjoint union yields the desired OR-composition, other problems seem to require involved constructions, for example \textsc{Set Cover} with parameter universe size~\cite{DomLS14}.
Indeed, devising a OR-composition can be quite challenging and the task becomes even harder when considering several, seemingly orthogonal parameterizations at once.

To illustrate the problem with such combined parameters, let us consider the NP-hard problem \spmveTsc{} (\spmveAcr{}).
Herein, an undirected graph $G=(V,E)$ with $s,t\in V$, and two integers $k,\ell\in\N$ are given, and the question is whether it is possible to delete at most~$k$ edges such that the shortest $s$-$t$~path is of length at least~$\ell$.
Using a simple branching algorithm, one can show that \spmveAcr{}($k,\ell$) is fixed-parameter tractable for the combined 
parameter~$(k,\ell)$~\cite{GolovachT11,BazganNN15}.
Golovach and Thilikos~\citet{GolovachT11} posed as an open problem whether \spmveAcr{}($k,\ell$) admits a polynomial kernel.
To exclude the existence of a polynomial kernel for \spmveAcr{}($k,\ell$), 
we would like to apply the OR-composition framework to the problem.%

A standard approach for applying the OR-composition technique to a problem like \spmveAcr{} would be to concatenate the input instances on the source and sink vertices, what one might refer to as ``serial composition'' (see, e.g., \cite{FluschnikKNS15,CyganKPPW14}). 
To this end, one needs some additional gadgets to ensure that only in one instance edges are deleted. 
This form of composition, however, induces a dependency of the second parameter~$\ell$ on the number of instances, which is not allowed. 
Another standard approach is introducing a global sink and source vertex, and connecting all source vertices with the global source and all sink vertices with the global sink, what one might refer to as a ``parallel composition''.
This form of composition would keep~$\ell$ small enough, but induces a dependency of the first parameter~$k$ on the number of instances. 
Summarizing, the parameter~$k$ seems to ask for a serial composition and the parameter~$\ell$ seems to ask for a parallel composition.
For some problems using a tree as ``instance selector'' was helpful, see for example Bevern et al.~\citet{BBCHHNS17} or Bazgan et al.~\citet{BCCFFL14}.
The problem with trees is that they introduce small (constant-size) $s$-$t$ cuts, which is problematic for \spmveTsc{}.

In this work, we introduce a fractal structure as an instance selector which has the nice properties of trees but does not introduce small cuts.
Our fractal structure allows avoiding the issues discussed above for serial and parallel compositions.
Thus, it can be used for excluding a polynomial kernel for \spmveAcr{}, as well as for other problems.
We mention that one can find the fractal structure in a proof by Guillemot et al.~\cite{GuillemotHPP13} refuting the existence of a polynomial kernel for a graph-coloring problem.

\subparagraph*{Our contributions.}
Our main technical contribution is to introduce a family of graphs that we call \fractal{}s and that build on triangles. 
\fractal{}s feature a fractal-like structure, in the sense of self-similarity and scale-invariance.
Using these \fractal{}s in OR-cross-compositions, we show that the following parameterized graph modification problems (definitions are stated subsequently) and several of their variants do not admit polynomial kernels (unless $\NP\subseteq \coNP/\poly$):
\begin{itemize}
 \item \spmveTsc{$(k,\ell)$} (\spmveAcr{$(k,\ell)$});%
 \item \dmveTsc{$(k,\ell)$} (\dmveAcr{$(k,\ell)$});%
 \item \gmveTsc{$(k,\ell)$} (\gmveAcr{$(k,\ell)$});%
\end{itemize}
\begin{table}
      \setlength{\tabcolsep}{10pt}
	\centering
	\begin{tabular}{lcccc}
		\toprule
		Problem & \multicolumn{4}{c}{edge deletion}  \\
		  & \multicolumn{2}{c}{directed} & \multicolumn{2}{c}{undirected} \\
		  & \multicolumn{2}{c}{planar/general} & planar & /general \\
		\midrule 
		\midrule
		\spmveAcr{}$(k,\ell)$ & \multicolumn{2}{c}{No PK [\Cref{thm:spmvenopolykernelforplanardag}/\Cref{thm:spmveondags}]} & \multicolumn{2}{c}{No PK [\Cref{thm:spmvenopolykernelforplanardag}/\Cref{theo:spmve}]} \\
		\dmveAcr{}$(k,\ell)$ & \multicolumn{2}{c}{No PK [\Cref{thm:dirmdednopk}]} & \multicolumn{2}{c}{No PK [\Cref{cor:dmvenopkplanar}/\Cref{thm:NoPolyKernel-D-MVE}]} \\
		\gmveAcr{}$(k,\ell)$ & \multicolumn{2}{c}{No PK [\Cref{thm:NoPolyKernelPlanarDirectedGMVE}/\Cref{thm:noppkforgmve}]} & PK~\cite{XiaZ11} & ? \\ 
		\bottomrule
	\end{tabular}
	\caption{Survey of the concrete results of this paper (under the assumption that $\NP\not\subseteq \coNP/\poly$). 
				PK stands for polynomial kernel and a ``?'' indicates that it is open whether a polynomial kernel exists. 
				We remark that the no-polynomial-kernel results for \spmveAcr{}$(k,\ell)$ on directed graphs still hold for directed acyclic graphs.
				Moreover, the results for the undirected variants also hold for the combined parameter $(k,\ell,\tw)$, where $\tw$ denotes the treewidth.
				Note that we claim without proof that, except for the planar variants, our proofs also transfer to the vertex deletion case, both for directed and undirected graphs.
	}
	\label{tab:overview}
\end{table}
\cref{tab:overview} surveys our no-polynomial-kernel results and spots an open question.

	The graph edge-modification problems~\spmveTsc{} (\spmveAcr{}), \dmveTsc{} (\dmveAcr{}), and \gmveTsc{} (\gmveAcr{}) are defined as follows. 
	The \spmveAcr{} problem asks, given an undirected graph~$G=(V,E)$, two vertices~$s,t\in V$, and two integers~$k,\ell$, whether there are at most~$k$~edge deletions such that the shortest $s$-$t$~path is of length at least~$\ell$. 
	The \dmveAcr{} problem asks, given an undirected connected graph~$G=(V,E)$ and two integers $k,\ell$, whether there are at most~$k$~edge deletions such that the remaining graph remains connected and has diameter at least~$\ell$. 
	The \gmveAcr{} problem asks, given a directed graph~$G=(V,E)$ and two integers~$k,\ell$, whether there are at most~$k$~edge deletions such that the remaining graph has no cycle of length smaller than~$\ell$.
	In addition, we consider several variants (planar, directed, vertex deletion) 
	of these problems.
	We remark that we also show that for the undirected (planar) variants, unless $\NP\not\subseteq \coNP/\poly$, \spmveAcr{} and \dmveAcr{} parameterized by~$(k,\ell,\tw)$ do not admit a polynomial kernel, where $\tw$ denotes the treewidth of the underlying graph~$G$.

\section{Preliminaries}
\label{sec:prelim}

	We use standard notation from parameterized complexity \cite{DowneyF13,FlumG06,Nie06,CyganFKLMPPS15} and graph theory \cite{Diestel10,West00}.
	Throughout this paper we denote by~$\log$ the logarithm with base two.
\subparagraph*{Graph Theory.}

Let~$G = (V, E)$ be a graph. 
	We denote by $V(G)$ the vertex set of~$G$ and by~$E(G)$ the edge set of~$G$. 
	For a vertex set $W \subseteq V (G)$ (edge set $F \subseteq E(G)$), we denote by~$G[W]$ ($G[F]$) the subgraph of~$G$ \emph{induced} by the vertex set~$W$ (edge set $F$).
For $C\subseteq V(G)$ ($C\subseteq E(G)$) we write $G-C$ for the graph~$G$ where all vertices (edges) in $C$ are deleted. 
	Note that the deletion of a vertex implies the deletion of all its incident edges.    

A cycle is a connected graph where every vertex has degree exactly two. 
The length of a cycle is the number of edges in the cycle. 
In directed graphs, a cycle is a connected graph where every vertex has outdegree and indegree exactly one. 
The \emph{girth} of a graph~$G$ is the length of the shortest cycle (contained) in~$G$.
	A tree is a simple, connected and cycle-free graph.
	A path is a tree with no vertex of degree at least three. 
	We call the vertices with degree one the endpoints of the path. 
	An $s$-$t$ path is a path where the vertices~$s$ and~$t$ are the endpoints of the path. 
	The length of a path is the number of edges in the path. 
	In directed graphs, an $s$-$t$ path is a path where all arcs are directed toward~$t$. 
The \emph{diameter} of a graph~$G$ is the maximum length of any shortest~$v$-$w$~path over all~$v, w\in V(G)$, $v\neq w$.

	Let~$G$~be an undirected, connected graph. 
	An edge~cut~$C\subseteq E(G)$ is a set of edges such that the graph $G- C$ is not connected. 
	Let $s, t \in V (G)$ be two vertices in~$G$. 
	An $s$-$t$ edge~cut~$C$ is an edge~cut such that the vertices $s$ and $t$ are not connected in~$G- C$. 
	If~$G$ is a directed graph, then $C\subseteq E(G)$ is an $s$-$t$~edge~cut if there is no $s$-$t$~path in~$G-C$. 
An $s$-$t$ edge~cut $C$ is called \emph{minimal} if $C\backslash \{e\}$ is not an $s$-$t$ edge~cut in~$G$ for all~$e\in C$. 
An $s$-$t$ edge~cut~$C$ is called \emph{minimum} if there is no $s$-$t$ edge~cut $C'$ in $G$ such that $|C'|<|C|$.

Given a graph $G=(V,E)$ and two non-adjacent vertices $v,w\in V$, we say we \emph{merge} the vertices $v$ and $w$ if we add a new vertex $vw$ to $V$ as well as the edge set $\{\{vw,x\}\mid \{x,v\}\in E\}\cup\{\{vw,x\}\mid \{x,w\}\in E\}$ to $E$, and we delete the vertices $v$ and $w$ and all incident edges to $v$ and $w$. 
\subparagraph*{Parameterized Complexity.}

A \emph{parameterized problem} is a set of instances $(\I,k)$ where $\I \in\Sigma^*$ for a finite alphabet $\Sigma$, and $k\in \mathbb{N}$ is the \emph{parameter}.
A parameterized problem~$L$ is \emph{fixed-parameter tractable (fpt)} if it can be decided in~$f(k)\cdot |\I|^{O(1)}$ time whether~$(\I,k)\in L$, where~$f$ is a computable function only depending on~$k$.
We say that two instances~$(\I,k)$ and $(\I',k')$ of parameterized problems~$P$ and~$P'$ are \emph{equivalent} if $(\I,k)$ is \yes for~$P$ if and only if $(\I',k')$ is \yes for~$P'$. 
A \emph{kernelization} is an algorithm that, given an instance~$(\I,k)$ of a parameterized problem~$P$, computes in polynomial time an equivalent instance~$(\I',k')$ of~$P$ (the \emph{kernel}) such that $|\I'|+k'\leq f(k)$ for some  computable function~$f$ only depending on~$k$. We say that $f$ measures the \emph{size} of the kernel, and if $f\in k^{O(1)}$, we say that $P$ admits a polynomial kernel. 
We remark that a decidable parameterized problem is fixed-parameter tractable if and only if it admits a kernel~\cite{CaiCDF97}.

In this paper we use the framework of Bodlaender et al.~\cite{BodlaenderJK14} which extends the notion of OR-composition to OR-cross-composition.
Given an \NP-hard problem~$L$, an equivalence relation~$\R$ on the instance of~$L$ is a \emph{polynomial equivalence relation} if 
\begin{enumerate}[(i)]
 \item one can decide for any two instances in time polynomial in their sizes whether they belong to the same equivalence class, and
 \item for any finite set~$S$ of instances, $\R$ partitions the set into at most~$(\max_{x \in S} |x|)^{O(1)}$ equivalence classes. %
\end{enumerate}
\begin{definition} 
	Given an \NP-hard problem~$L$, a parameterized problem $P$, and a polynomial equivalence relation~$\R$ on the instances of $L$, an \emph{OR-cross-composition} of $L$ into $P$ (with respect to $\R$) is an algorithm that takes $\ell$ $\R$-equivalent instances $\I_1,\ldots,\I_\ell$ of $L$ and constructs in time polynomial in $\sum_{i=1}^\ell |\I_\ell|$ an instance $(\I,k)$ such that 
	\begin{compactenum}
		\item $k$ is polynomially upper-bounded in $\max_{1\leq i\leq \ell}|\I_i|+\log(\ell)$ and 
		\item $(\I,k)$ is \yes for $P$ if and only if there is at least one~$\ell'\in[\ell]$ such that $\I_{\ell'}$ is \yes for $L$. 
	\end{compactenum}
\end{definition}
If a parameterized problem~$P$ admits an OR-cross-composition for some \NP-hard problem~$L$, then~$P$ does not admit a polynomial kernel with respect to its parameterization, unless $\NP\subseteq \coNP/\poly$ \cite{BodlaenderJK14}. 
We remark that we can assume that $\ell=2^j$ for some $j\in\N$ since we can add trivial \no-instances from the same equivalence class to reach a power of two. 
We refer to the survey of~Kratsch~\citet{Kratsch14} for an overview on kernelization and lower bounds.

\section{The ``Fractalism'' Technique}

In this section, we describe our new technique based on~\emph{triangle fractals} (\emph{\fractal{s}} for short). 
We provide a general construction scheme for cross-com\-po\-si\-tions using \fractal{s}. 
To this end, we first define \fractal{s} and then discuss several of their properties in \cref{subsec:prop}. 
Furthermore, we present in \cref{ssec:directedCase} a directed variant and provide two ``construction manuals'' for an application of~\fractal{s} in cross-compositions in \cref{ssec:apps}.

\begin{figure}[t]
	\centering
	\begin{tikzpicture}%
\tikzstyle{tnode}=[fill, circle, scale=1/2, draw];
\tikzstyle{tnodespez}=[circle, scale=1/2, draw];

\node (d1) at (0,0)[tnodespez,label=270:{$\lv$}]{};
\node (d2) at (2,0)[tnodespez,label=270:{$\rv$}]{};
\node (d3) at (1,1)[tnode]{};
\draw (d1) -- (d2) -- (d3) --(d1);

\def\x{3.5}

\node (d1) at (0+\x,0)[tnodespez,label=270:{$\lv$}]{};
\node (d2) at (2+\x,0)[tnodespez,label=270:{$\rv$}]{};
\node (d3) at (1+\x,1)[tnode]{};
\draw (d1) -- (d2) -- (d3) --(d1);

\node (d21) at (-0.5+\x,1.5)[tnode]{};
\node (d22) at (2+0.5+\x,1.5)[tnode]{};

\draw (d1) -- (d21) -- (d3);
\draw (d2) -- (d22) -- (d3);

\def\xx{8}

\node (d1) at (0+\xx,0)[tnodespez,label=270:{$\lv$}]{};
\node (d2) at (2+\xx,0)[tnodespez,label=270:{$\rv$}]{};
\node (d3) at (1+\xx,1)[tnode]{};
\draw (d1) -- (d2) -- (d3) --(d1);

\node (d21) at (-0.5+\xx,1.5)[tnode]{};
\node (d22) at (2+0.5+\xx,1.5)[tnode]{};

\draw (d1) -- (d21) -- (d3);
\draw (d2) -- (d22) -- (d3);

\node (d31) at (-1.5+\xx,0.75)[tnode]{};
\node (d32) at (0.25+\xx,2.25)[tnode]{};
\node (d33) at (2-0.25+\xx,2.25)[tnode]{};
\node (d34) at (2+1.5+\xx,0.75)[tnode]{};

\draw (d1) -- (d31) -- (d21);
\draw (d3) -- (d32) -- (d21);
\draw (d3) -- (d33) -- (d22);
\draw (d2) -- (d34) -- (d22);

\def\xxx{5}
\def\y{-4}

\node (d1) at (0+\xxx,0+\y)[tnodespez,label=270:{$\lv$}]{};
\node (d2) at (2+\xxx,0+\y)[tnodespez,label=270:{$\rv$}]{};
\node (d3) at (1+\xxx,1+\y)[tnode]{};
\draw (d1) -- (d2) -- (d3) --(d1);

\node (d21) at (-0.5+\xxx,1.5+\y)[tnode]{};
\node (d22) at (2+0.5+\xxx,1.5+\y)[tnode]{};

\draw (d1) -- (d21) -- (d3);
\draw (d2) -- (d22) -- (d3);

\node (d31) at (-1.5+\xxx,0.75+\y)[tnode]{};
\node (d32) at (0.25+\xxx,2.25+\y)[tnode]{};
\node (d33) at (2-0.25+\xxx,2.25+\y)[tnode]{};
\node (d34) at (2+1.5+\xxx,0.75+\y)[tnode]{};

\draw (d1) -- (d31) -- (d21);
\draw (d3) -- (d32) -- (d21);
\draw (d3) -- (d33) -- (d22);
\draw (d2) -- (d34) -- (d22);

\node (d41) at (-1+\xxx,0.1+\y)[tnode]{};
\node (d42) at (-1.25+\xxx,1.75+\y)[tnode]{};
\node (d43) at (-0.5+\xxx,2.5+\y)[tnode]{};
\node (d44) at (0.75+\xxx,2.75+\y)[tnode]{};
\node (d48) at (2+1+\xxx,0.1+\y)[tnode]{};
\node (d47) at (2+1.25+\xxx,1.75+\y)[tnode]{};
\node (d46) at (2+0.5+\xxx,2.5+\y)[tnode]{};
\node (d45) at (2-0.75+\xxx,2.75+\y)[tnode]{};

\draw (d1) -- (d41) -- (d31);
\draw (d21) -- (d42) -- (d31);
\draw (d21) -- (d43) -- (d32);
\draw (d3) -- (d44) -- (d32);
\draw (d3) -- (d45) -- (d33);
\draw (d22) -- (d46) -- (d33);
\draw (d22) -- (d47) -- (d34);
\draw (d2) -- (d48) -- (d34);

\end{tikzpicture}
	\caption{\fractal{}s $\triangle_1,\triangle_2,\triangle_3,\triangle_4$. The two special vertices~$\lv$ and~$\rv$ are highlighted by empty circles.}
	\label{fig:trifrac24816}
\end{figure}
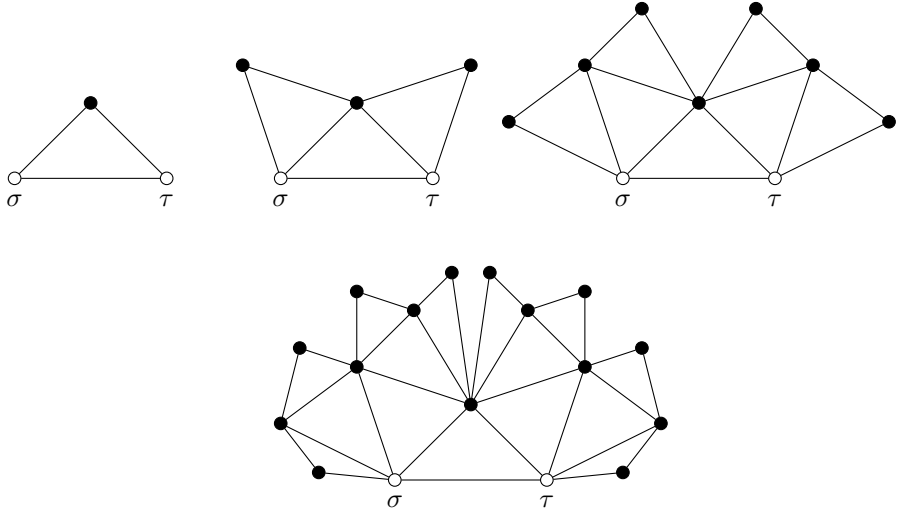

Roughly speaking, a \fractal{} can be constructed by iteratively putting triangles on top of each other, see \cref{fig:trifrac24816} for four~examples.

\begin{definition}\label[definition]{def:fractal}
	For $q\geq 1$, the \emph{$q$-\fractal}~$\triangle_q$ is the graph constructed as follows:
	\begin{compactenum}[(1)]
		\item Set $\triangle_0:= \{\lv,\rv\}$ with $\{\lv,\rv\}$ being a ``marked edge'' with endpoints~$\lv$ and~$\rv$, subsequently referred to as special vertices.
		\item Let~$F$ be the set of marked edges.
		\item For each edge $e\in F$, add a new vertex and connect it by two new edges with the endpoints of~$e$, and mark the two added edges.%
		\item Unmark all edges in $F$.
		\item Repeat (2)-(4) $q-1$ times.
	\end{compactenum}
\end{definition}
The fractal structure of~$\triangle_q$ might be easier to see when considering the following equivalent recursive definition of~$\triangle_q$:
For the base case we define~$\triangle_0:= \{\lv,\rv\}$ as in \cref{def:fractal}.
Then, the $q$-\fractal~$\triangle_q$ is constructed as follows. 
Take two $(q-1)$-\fractal{s} $\triangle_{q-1}'$ and $\triangle_{q-1}''$, where $\lv',\rv'$ and $\lv'',\rv''$ are the special vertices of~$\triangle_{q-1}'$ and~$\triangle_{q-1}''$, respectively. 
Then $\triangle_q$ is obtained by merging the vertices $\rv'$ and $\lv''$, adding the edge $\{\lv',\rv''\}$, and setting $\lv=\lv'$ and $\rv=\rv''$ as the special vertices of $\triangle_q$.
We remark that we make use of the recursive structure in later proofs.
However, by construction, we immediately obtain the following (for the latter, see e.g.~\cite{Bodlaender98}).
\begin{observation}\label{obs!outerplanar}
 The \fractal{} is outerplanar and hence the treewidth of $\triangle_q$ is~$\tw(\triangle_q)\leq 2$ for every $q\in \N$.
\end{observation}

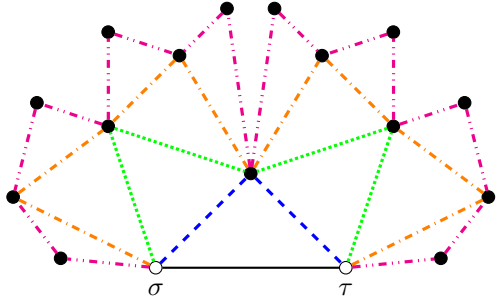
\begin{figure}[t]
	\centering
	\begin{tikzpicture}[xscale=1.25, yscale=1.25]

\tikzstyle{tnode}=[fill, circle, scale=1/2, draw];
\tikzstyle{tnodespez}=[circle, scale=1/2, draw];

\def\xx{6.5}
\def\y{0}

\node (d1) at (0+\xx,0+\y)[tnodespez,label=270:{$\lv$}]{};
\node (d2) at (2+\xx,0+\y)[tnodespez,label=270:{$\rv$}]{};
\node (d3) at (1+\xx,1+\y)[tnode]{};
\draw[thick] (d1) -- (d2);
\draw[color=blue, dashed, very thick] (d1) --(d3) --(d2);

\node (d21) at (-0.5+\xx,1.5+\y)[tnode]{};
\node (d22) at (2+0.5+\xx,1.5+\y)[tnode]{};

\draw[color=green, densely dotted, very thick] (d1) -- (d21) -- (d3);
\draw[color=green, densely dotted, very thick] (d2) -- (d22) -- (d3);

\node (d31) at (-1.5+\xx,0.75+\y)[tnode]{};
\node (d32) at (0.25+\xx,2.25+\y)[tnode]{};
\node (d33) at (2-0.25+\xx,2.25+\y)[tnode]{};
\node (d34) at (2+1.5+\xx,0.75+\y)[tnode]{};

\draw[color=orange, dashdotted, very thick] (d1) -- (d31) -- (d21);
\draw[color=orange, dashdotted, very thick] (d3) -- (d32) -- (d21);
\draw[color=orange, dashdotted, very thick] (d3) -- (d33) -- (d22);
\draw[color=orange, dashdotted, very thick] (d2) -- (d34) -- (d22);

\node (d41) at (-1+\xx,0.1+\y)[tnode]{};
\node (d42) at (-1.25+\xx,1.75+\y)[tnode]{};
\node (d43) at (-0.5+\xx,2.5+\y)[tnode]{};
\node (d44) at (0.75+\xx,2.75+\y)[tnode]{};
\node (d48) at (2+1+\xx,0.1+\y)[tnode]{};
\node (d47) at (2+1.25+\xx,1.75+\y)[tnode]{};
\node (d46) at (2+0.5+\xx,2.5+\y)[tnode]{};
\node (d45) at (2-0.75+\xx,2.75+\y)[tnode]{};

\draw[color=magenta, dashdotdotted, very thick] (d1) -- (d41) -- (d31);
\draw[color=magenta, dashdotdotted, very thick] (d21) -- (d42) -- (d31);
\draw[color=magenta, dashdotdotted, very thick] (d21) -- (d43) -- (d32);
\draw[color=magenta, dashdotdotted, very thick] (d3) -- (d44) -- (d32);
\draw[color=magenta, dashdotdotted, very thick] (d3) -- (d45) -- (d33);
\draw[color=magenta, dashdotdotted, very thick] (d22) -- (d46) -- (d33);
\draw[color=magenta, dashdotdotted, very thick] (d22) -- (d47) -- (d34);
\draw[color=magenta, dashdotdotted, very thick] (d2) -- (d48) -- (d34);

\end{tikzpicture}
	\caption{Highlighting the different boundaries of~$\triangle_4$ by line-types (solid: boundary~$B_0$; dashed: boundary~$B_1$; dotted: boundary~$B_2$; dash-dotted: boundary~$B_3$; dash-dot-dotted: boundary~$B_4$).}
	\label{fig:trifrac16}
\end{figure}

In the $i$th execution of (2)-(4) in \cref{def:fractal}, we obtain $2^{i-1}$ new triangles. 
We say that these triangles have depth $i$. 
The \emph{boundary}~$B_i\subseteq E(\triangle_q)$, $i\in[q]$, are those edges of the triangles of depth $i$ which are not edges of the triangles of depth $i-1$. 
As a convention, the edge~$\{\lv,\rv\}$ connecting the two special vertices~$\lv$ and~$\rv$ forms the boundary~$B_0$. 
We refer to~\cref{fig:trifrac16} for an illustration of the boundaries in the \fractal{}~$\triangle_4$. 
Moreover, by construction, we obtain the following:
\begin{observation}\label{obs:boundaries}
	In every~\fractal{}, each boundary forms a $\lv$-$\rv$~path, and all boundaries are pairwise edge-disjoint. 
\end{observation} 
Note that the boundary~$B_q$ contains $p=2^q$~edges. 
Thus, the number of edges in~$\triangle_q$ is~$\sum_{i=0}^{q} 2^i = 2^{q+1} - 1 = 2\cdot p-1$. 
Further observe that all vertices of~$\triangle_q$ are incident with the edges in~$B_q$, and $B_q$ forms a~$\lv$-$\rv$~path. 
Hence, $\triangle_q$ contains $p+1$~vertices. 

\subparagraph*{Reducing the Weighted to the Unweighted Case.}
The \emph{weighted} \fractal{} is the \fractal{} equipped with \emph{edge~costs}, that is, the cost for deleting an edge in the \fractal{}. 
If all edges in~$\triangle_q$ are of the same edge~cost~$c\in\N$, then we write $\triangle_q^c$ (we drop the superscript if~$c=1$).
In the remainder of the paper, we focus on the unweighted case of \fractal{s} without multiple edges or loops. 
This is possible due to the following reduction of the weighted to the unweighted case. 
Consider the weighted~\fractal{}~$\triangle_q^c$ with $c\geq 2$. 
To reduce to the case with an unweighted, simple graph, we add $c-1$ further copies for each edge. 
Thus, to make two adjacent vertices non-adjacent, it requires~$c$ edge-deletions. 
To make the graph simple, we subdivide each edge. 
We remark that in this way we double the distances of the vertices in the original \fractal{}. 
Thus, whenever we consider distances in the fractal with edge~cost and the graph obtained by the reduction above, we have to take into account a factor of two. 

Finally, let us remark that the treewidth of the graph $G$ obtained by the modification of the \fractal{} described above remains at most two, though it is not necessarily outerplanar anymore.
To see this, observe that outerplanar graphs are series-parallel.
Moreover, a graph obtained by replacing an edge in a series-parallel graph by a number of paths with the same endpoints remains series-parallel.
It follows that $G$ has treewidth at most two.

\subsection{Properties of~\Fractal{s}}\label{subsec:prop}

The goal of this subsection is to prove several properties of~\fractal{s} that are used in later constructions. 
Some key properties of~\fractal{s} appear in the context of $\lv$-$\rv$~edge~cuts in~$\triangle_q$. 
To prove other properties, we later introduce the notion of the dual structure behind the \fractal{}s.

The minimum edge~cuts in~$\triangle_q$ will play a central role when using \fractal{s} in cross-compositions since the minimum edge~cuts serve as instance selectors (see~\cref{ssec:apps}). 
First, we discuss the size and structure of the minimum edge~cuts in~$\triangle_q$. 

\begin{lemma}\label[lemma]{lem:MinCutOfSizeqPlusOne}
	Every minimum $\lv$-$\rv$~edge~cut in~$\triangle_q$ is of size~$q+1$. 
\end{lemma}
\begin{proof}
	Let $C$ be a minimum $\lv$-$\rv$~edge~cut in~$\triangle_q$. Note that the degrees of~$\lv$ and~$\rv$ are exactly~$q+1$, and thus~$|C|\leq q+1$. 
	Moreover, the boundaries in~$\triangle_q$ are pairwise edge-disjoint and each boundary forms a $\lv$-$\rv$~path (\cref{obs:boundaries}). 
	Since~$\triangle_q$ contains $q+1$~boundaries, it follows that there are at least $q+1$~disjoint $\lv$-$\rv$~paths in~$\triangle_q$. 
	Menger's theorem~\cite{menger1927} states that in a graph with distinct source and sink, the maximum number of disjoint source-sink paths equals the minimum size of any source-sink edge cut.
	Thus, by Menger's theorem, it follows that~$|C|\geq q+1$. 
	Hence~$|C|=q+1$.
\end{proof}

From the fact that the boundaries are pairwise edge-disjoint and each boundary forms a $\lv$-$\rv$~path, we can immediately derive the following from~\cref{lem:MinCutOfSizeqPlusOne}.

\begin{corollary}\label{cor:MinCutBoundaryEdges}
Every minimum $\lv$-$\rv$~edge~cut in~$\triangle_q$ contains exactly one edge of each boundary. 
\end{corollary}
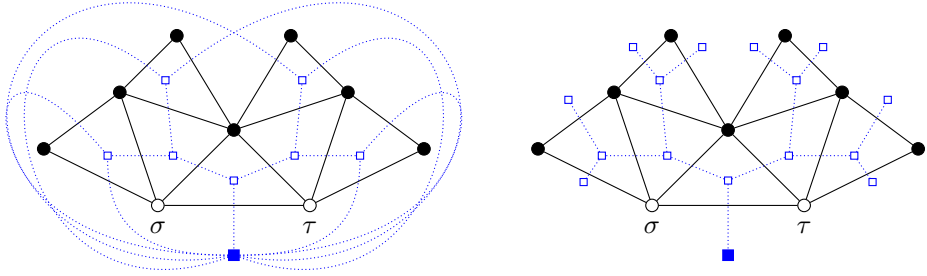
\begin{figure}[t]
	\centering
	\begin{tikzpicture}[auto]

\tikzstyle{tnode}=[fill, circle, scale=1/2, draw];
\tikzstyle{tnodespez}=[circle, scale=1/2, draw];
\tikzstyle{tnodedual}=[color=blue, scale=0.4, draw];

\node (bbu) at (0,2.7)[]{};
\node (bbu) at (0,-0.8)[]{};
\node (bbu) at (-1.9,0)[]{};

\def\x{0}
\def\y{0}

\node (d1) at (0+\x,0)[tnodespez,label=270:{$\lv$}]{};
\node (d2) at (2+\x,0)[tnodespez,label=270:{$\rv$}]{};
\node (d3) at (1+\x,1)[tnode]{};
\draw (d1) -- (d2) -- (d3) --(d1);

\node (d21) at (-0.5+\x,1.5)[tnode]{};
\node (d22) at (2+0.5+\x,1.5)[tnode]{};

\draw (d1) -- (d21) -- (d3);
\draw (d2) -- (d22) -- (d3);

\node (d31) at (-1.5+\x,0.75)[tnode]{};
\node (d32) at (0.25+\x,2.25)[tnode]{};
\node (d33) at (2-0.25+\x,2.25)[tnode]{};
\node (d34) at (2+1.5+\x,0.75)[tnode]{};

\draw (d1) -- (d31) -- (d21);
\draw (d3) -- (d32) -- (d21);
\draw (d3) -- (d33) -- (d22);
\draw (d2) -- (d34) -- (d22);

\node (p1) at (1+\x,0.33)[tnodedual]{};

\node (p12) at (2-0.2+\x,0.66)[tnodedual]{};
\node (p11) at (0.2+\x,0.66)[tnodedual]{};
\draw [color=blue, densely dotted] (p1) -- (p11);
\draw [color=blue, densely dotted] (p1) -- (p12);

\node (p21) at (\x-0.66,0.66)[tnodedual]{};
\node (p22) at (\x+0.1,1.66)[tnodedual]{};
\node (p24) at (\x+2+0.66,0.66)[tnodedual]{};
\node (p23) at (\x+2-0.1,1.66)[tnodedual]{};
\draw[color=blue, densely dotted] (p11) -- (p21);
\draw[color=blue, densely dotted] (p11) -- (p22);
\draw[color=blue, densely dotted] (p12) -- (p23);
\draw[color=blue, densely dotted] (p12) -- (p24);
\node (p0) at (1+\x,-0.33*2)[scale=1.5,tnodedual,fill]{};
\draw[color=blue, densely dotted] (p1) -- (p0);
\begin{pgfinterruptboundingbox}
\draw[color=blue, densely dotted,-] (p0) to [out=180, in=270, looseness=1.25](p21);
\draw[color=blue, densely dotted,-] (p0) to [out=190, in=130, looseness=4](p21);
\draw[color=blue, densely dotted,-] (p0) to [out=200, in=140, looseness=3.5](p22);
\draw[color=blue, densely dotted,-] (p0) to [out=0, in=40, looseness=5](p22);
\draw[color=blue, densely dotted,-] (p0) to [out=0, in=270, looseness=1.25](p24);
\draw[color=blue, densely dotted,-] (p0) to [out=350, in=50, looseness=4](p24);
\draw[color=blue, densely dotted,-] (p0) to [out=340, in=40, looseness=3.5](p23);
\draw[color=blue, densely dotted,-] (p0) to [out=180, in=140, looseness=5](p23);
\end{pgfinterruptboundingbox}

\def\x{6.5}
\node (d1) at (0+\x,0)[tnodespez,label=270:{$\lv$}]{};
\node (d2) at (2+\x,0)[tnodespez,label=270:{$\rv$}]{};
\node (d3) at (1+\x,1)[tnode]{};
\draw (d1) -- (d2) -- (d3) --(d1);

\node (d21) at (-0.5+\x,1.5)[tnode]{};
\node (d22) at (2+0.5+\x,1.5)[tnode]{};

\draw (d1) -- (d21) -- (d3);
\draw (d2) -- (d22) -- (d3);

\node (d31) at (-1.5+\x,0.75)[tnode]{};
\node (d32) at (0.25+\x,2.25)[tnode]{};
\node (d33) at (2-0.25+\x,2.25)[tnode]{};
\node (d34) at (2+1.5+\x,0.75)[tnode]{};

\draw (d1) -- (d31) -- (d21);
\draw (d3) -- (d32) -- (d21);
\draw (d3) -- (d33) -- (d22);
\draw (d2) -- (d34) -- (d22);

\node (p1) at (1+\x,0.33)[tnodedual]{};

\node (p12) at (2-0.2+\x,0.66)[tnodedual]{};
\node (p11) at (0.2+\x,0.66)[tnodedual]{};
\draw [color=blue, densely dotted] (p1) -- (p11);
\draw [color=blue, densely dotted] (p1) -- (p12);

\node (p21) at (\x-0.66,0.66)[tnodedual]{};
\node (p22) at (\x+0.1,1.66)[tnodedual]{};
\node (p24) at (\x+2+0.66,0.66)[tnodedual]{};
\node (p23) at (\x+2-0.1,1.66)[tnodedual]{};
\draw[color=blue, densely dotted] (p11) -- (p21);
\draw[color=blue, densely dotted] (p11) -- (p22);
\draw[color=blue, densely dotted] (p12) -- (p23);
\draw[color=blue, densely dotted] (p12) -- (p24);

\node (p31) at (\x-0.9,0.31)[tnodedual]{};
\node (p32) at (\x-1.1,1.4)[tnodedual]{};
\node (p33) at (\x-0.25,2.1)[tnodedual]{};
\node (p34) at (\x+0.66,2.1)[tnodedual]{};
\node (p38) at (\x+2+0.9,0.31)[tnodedual]{};
\node (p37) at (\x+2+1.1,1.4)[tnodedual]{};
\node (p36) at (\x+2+0.25,2.1)[tnodedual]{};
\node (p35) at (\x+2-0.66,2.1)[tnodedual]{};

\draw[color=blue, densely dotted] (p21) -- (p31);
\draw[color=blue, densely dotted] (p21) -- (p32);
\draw[color=blue, densely dotted] (p22) -- (p33);
\draw[color=blue, densely dotted] (p22) -- (p34);
\draw[color=blue, densely dotted] (p23) -- (p35);
\draw[color=blue, densely dotted] (p23) -- (p36);
\draw[color=blue, densely dotted] (p24) -- (p37);
\draw[color=blue, densely dotted] (p24) -- (p38);

\node (p0) at (1+\x,-0.33*2)[scale=1.5,tnodedual,fill]{};
\draw[color=blue, densely dotted] (p1) -- (p0);
 
\end{tikzpicture}
	\caption{Left: The \fractal{}~$\triangle_3$ (circles and solid lines) and its dual graph (squares and dotted lines). The filled square is the vertex dual to the outer face in the dual graph. Right: The \fractal{}~$\triangle_3$ (circles and solid lines) and its dual structure~$T_3$, illustrated by squares and dotted lines, where the filled square corresponds to the root of the dual structure.}
	\label{fig:trifrac16dual}
\end{figure}

In the following we describe a (hidden) dual structure in~$\triangle_q$, that is, a complete binary tree with $p$~leaves. 
We refer to~\cref{fig:trifrac16dual} for an example of the dual structure in~$\triangle_3$. 
To talk about the dual structure by means of duality of plane graphs, %
we need a plane embedding of~$\triangle_q$. 
Hence we assume that~$\triangle_q$ is embedded as in~\cref{fig:trifrac24816} (iteratively extended). 
By~$T_q$ we denote the dual structure in~$\triangle_q$, where the vertex dual to the outer face is replaced by $p+1$ vertices (\emph{split vertices}) such that each edge incident with the dual vertex is incident with exactly one split vertex.
We consider the split vertex incident with the vertex dual to the triangle containing the edge~$\{\lv,\rv\}$ as the root vertex of the dual structure~$T_q$. 
Thus, the other split vertices correspond to the leaves of the dual structure~$T_q$. 
Note that the depth of a triangle one-to-one corresponds to the depth of the dual vertex in~$T_q$. 

Observe that there is a one-to-one correspondence between the edges in~$T_q$ and the edges in~$\triangle_q$. 
The following lemma states duality of root-leaf paths in~$T_q$ and minimum $\lv$-$\rv$~edge~cuts in $\triangle_q$, demonstrating the utility of the dual structure~$T_q$.

\begin{lemma}\label[lemma]{lem:path1to1cut}
	There is a one-to-one correspondence between root-leaf~paths in the dual structure~$T_q$ of~$\triangle_q$ and minimum $\lv$-$\rv$~edge~cuts in~$\triangle_q$. 
	Moreover, there are exactly $p=2^q$ pairwise different minimum $\lv$-$\rv$~edge~cuts in~$\triangle_q$. 
\end{lemma}
\begin{proof}
 Observe that each path from the root to a leaf in the dual structure~$T_q$ corresponds to a cycle in the dual graph. 
 It is well-known that there is a one-to-one correspondence between minimal edge~cuts in a plane graph and cycles in its dual graph~\cite[Proposition 4.6.1]{Diestel10}. 
 Herein, every cycle in the dual graph that ``cuts'' the edge~$\{\lv,\rv\}$ in~$\triangle_q$ is a root-leaf path in~$T_q$. 
 Thus, the only minimal $\lv$-$\rv$~edge~cuts are those corresponding to the root-leaf paths. 
 By the one-to-one correspondence of the depth of the triangles in~$\triangle_q$ and the depth of the vertices in~$T_q$, these edge~cuts are of cardinality~$q+1$.
 Hence, by~\cref{lem:MinCutOfSizeqPlusOne}, these edge~cuts are minimum edge~cuts. 
 
 Since $|B_q|=p$, there are exactly $p$ leaves in~$T_q$, and thus there are exactly $p$~different root-leaf paths in~$T_q$. 
 It follows that the number of pairwise different minimum~$\lv$-$\rv$~edge~cuts in~$\triangle_q$ is exactly $p=2^q$.
\end{proof}

Further, we obtain the following.

\begin{lemma}\label[lemma]{lem:DistToCompEnd}
	Let~$C$ be a minimum $\lv$-$\rv$~edge~cut in $\triangle_q$. 
	Let $\{x,y\}=C\cap B_q$, where $x$ is in the same connected component as $\lv$ in $\triangle_q- C$. 
	Then $\dist(\lv,x)+\dist(y,\rv)=q$ in~$\triangle_q- C$. 
\end{lemma}
\begin{proof}
	We prove the lemma by induction on~$q$.
	For the base case~$q=0$, observe that~$C = \{\lv,\rv\}$ and $\dist_{\triangle_0- C}(\lv,x)+\dist_{\triangle_0- C}(y,\rv)=0$.

	For the induction step, assume that the statement of the lemma is true for~$\triangle_{q-1}$.
	Now, let~$C$ be a minimum $\lv$-$\rv$~edge~cut in $\triangle_q$. 
	Hence, $\{\lv,\rv\} \in C$.
	Denote by~$u$ the (unique) vertex that is adjacent to the two special vertices~$\lv$ and~$\rv$.
	Let~$\triangle_{q-1}'$ and~$\triangle_{q-1}''$ be the two $(q-1)$-T-subfractals of~$\triangle_q$, so that~$\triangle_{q-1}'$ ($\triangle_{q-1}''$) has the special vertices~$\lv$ and~$u$ ($u$~and $\rv$).
	By~\cref{lem:path1to1cut}, the minimum $\lv$-$\rv$~edge~cut~$C$ corresponds to a root-leaf path in~$T_q$. 
	Hence, $C' := C \setminus \{\lv,\rv\}$ is either a subset of~$E(\triangle_{q-1}')$ or of~$E(\triangle_{q-1}'')$.
	Assume w.l.o.g.\ that~$C' \subseteq E(\triangle_{q-1}')$.
	It follows from the induction hypothesis that~$\dist_{\triangle_{q-1}'- C'}(\lv,x)+\dist_{\triangle_{q-1}'- C'}(y,u)=q-1$.
	Since~$\dist_{\triangle_{q}- C}(y,\rv) = \dist_{\triangle_{q-1}'- C'}(y,u) + 1$, it follows that $\dist_{\triangle_q- C}(\lv,x)+\dist_{\triangle_q- C}(y,\rv)=q$.
\end{proof}

\begin{remark}
 By an inductive proof like the one of~\cref{lem:DistToCompEnd}, one can easily show that the maximum degree $\Delta$ of $\triangle_q$ is exactly $2\cdot q$ for $q>0$. 
 Moreover, due to~\cref{lem:DistToCompEnd}, the diameter of $\triangle_q$ is bounded in~$O(q)$.
\end{remark}

Another observation on $\triangle_q$ is that any deletion of $d$~edges increases the length of any shortest $\lv$-$\rv$~path to at most $d+1$, unless the edge~deletion forms a $\lv$-$\rv$~edge~cut.

\begin{lemma}\label[lemma]{lem:lengthofpathsbyone}
	Let $D\subseteq E(\triangle_q)$ be a subset of edges of~$\triangle_q$. 
	If $D$ is not a $\lv$-$\rv$~edge~cut, then there is a $\lv$-$\rv$~path of length at most~$|D|+1$ in $\triangle_q- D$. 
\end{lemma}
\begin{proof}
We prove the statement of the lemma by induction on~$q$.
For the induction base with~$q=0$, observe that since~$D$ is not a $\lv$-$\rv$~edge~cut, it follows that~$D = \emptyset$ and, hence, $\lv$ and~$\rv$ have distance one.

For the induction step, assume that the statement of the lemma is true for~$\triangle_{q-1}$.
Now, let~$D\subseteq E(\triangle_q)$ be a subset of edges of~$\triangle_q$ such that~$D$ is not a $\lv$-$\rv$~edge~cut.
If~$\{\lv,\rv\} \notin D$, then there is a $\lv$-$\rv$~path of length one and the statement of the lemma holds.
Now consider the case $\{\lv,\rv\} \in D$.
Denote by~$u$ the (unique) vertex that is adjacent to the two special vertices~$\lv$ and~$\rv$.
If $\{\lv,\rv\} \in D$, then every $\lv$-$\rv$~path in $\triangle_q- D$ contains~$u$ and hence $\dist_{\triangle_q- D}(\lv,\rv) = \dist_{\triangle_q- D}(\lv,u) + \dist_{\triangle_q- D}(u,\rv)$.
(If there is no $\lv$-$u$-path or no $u$-$\rv$-path in $\triangle_q- D$, then~$D$ is a $\lv$-$\rv$~edge~cut; a contradiction to the assumption of the lemma.)
Now let~$\triangle_{q-1}'$ and~$\triangle_{q-1}''$ be the two $(q-1)$-T-subfractals of~$\triangle_q$, so that~$\triangle_{q-1}'$ ($\triangle_{q-1}''$) has the special vertices~$\lv$ and~$u$ ($u$ and $\rv$).
It follows that~$D$ can be partitioned into~$D = D' \cup D'' \cup \{\lv,\rv\}$ with~$D' \subseteq E(\triangle_{q-1}')$ and~$D'' \subseteq E(\triangle_{q-1}'')$.
By induction hypothesis, it follows that there is a $\lv$-$u$~path of length at most~$|D'|+1$ in $\triangle_{q-1}'- D'$ and a $u$-$\rv$~path of length at most~$|D''|+1$ in $\triangle_{q-1}''- D''$. 
Hence, there is a $\lv$-$\rv$~path of length at most~$|D'|+|D''|+2 = |D|+1$ in $\triangle_q- D$. 
\end{proof}
By~\cref{lem:lengthofpathsbyone}, the distance of the two special vertices $\lv$ and $\rv$ is upper-bounded by the number of edge deletions, where the deleted edges do no form a $\lv$-$\rv$ edge~cut. 
Hence, if only few edges are deleted in~$\triangle_q$, then~$\lv$ and~$\rv$ are not far away from each other.
The next lemma generalizes this by stating that the distance of \emph{any} vertex in~$\triangle_q$ to~$\lv$ or to~$\rv$ is quite small, even if a few edges are deleted.
Here ``quite small'' means that if~$O(q)$ edges are deleted, then the distance is still~$O(q)$ which is logarithmic in the size of~$\triangle_q$.
\begin{lemma}\label[lemma]{lem:lengthofanyvertex}
	Let $D\subseteq E(\triangle_q)$ be a subset of edges of~$\triangle_q$ and let~$x$ be an arbitrary vertex in~$V(\triangle)$.
	\begin{enumerate}[(A)]
		\item If $\triangle_q- D$ is connected, then $\dist_{\triangle_q- D}(\lv,x)\leq q + |D| + 1$ for all $x\in V(\triangle_q)$. 
		\item If $\triangle_q- D$ has exactly two connected components, with $\lv$ and $\rv$ being in different components, then $\min_{z\in\{\lv,\rv\}}\{\dist_{\triangle_q- D}(z,x)\}\leq q + |D| - 1$ for all $x\in V(\triangle_q)$. 
	\end{enumerate}
\end{lemma}

\begin{proof}
	We prove the two statements (A) and (B) simultaneously with an induction on depth~$q$ of the~\fractal{}.
 
	The base case is~$q=0$.
	For statement~(A), observe that $D=\emptyset$. 
	Thus, since $\rv$ has distance one to $\lv$, statement (A) follows. 
	For statement~(B), observe that $D=\{\{\lv,\rv\}\}$. 
	Thus, statement (B) holds.
 
	As our induction hypothesis, we assume that (A) and (B) hold for $1,\ldots, q-1$. 
	We write IH.(A) and IH.(B) for the induction hypothesis of (A) and (B), respectively. 
	We introduce some notation used for the induction step for both statements.
	Let~$\triangle_q$, $q > 0$, the \fractal with special vertices~$\lv$ and~$\rv$ and let~$u$ be the (unique) vertex in~$\triangle_q$ that is adjacent to~$\lv$ and~$\rv$, that is, $u$ is on the boundary~$B_1$ of~$\triangle_q$.
	Denote with~$\triangle_{q-1}'$ ($\triangle_{q-1}''$) the left (right) subfractal of~$\triangle_q$ with special vertices~$\lv$ and~$u$ ($u$ and~$\rv$).
	Furthermore, let~$D'$ ($D''$) be the subset of edges of $D$ deleted in $\triangle_{q-1}'$ ($\triangle_{q-1}''$).

	For the inductive step, we consider the two cases~$\{\lv,\rv\}\not\in D$ and~$\{\lv,\rv\} \in D$.
 
	\textbf{Case 1: $\{\lv,\rv\}\not\in D$.} 
	Obviously, this case excludes (B), since $\lv$ and $\rv$ are in the same connected component. 
	Thus, we consider the induction step for~(A). 
	Let $x$ be in the left subfractal~$\triangle_{q-1}'$.
	If $D'$ does not form an edge~cut in~$\triangle_{q-1}'$, then by IH.(A) it follows that $\dist_{\triangle_{q-1}'- D'}(\lv,x)\leq q-1+|D'|+1 \le q + |D|$. 
	Thus, we consider the case where $D'$ forms an edge~cut in~$\triangle_{q-1}'$. 
	Observe that such an edge~cut fulfills the requirements of statement (B) for~$\triangle_{q-1}'$. 
	By IH.(B), it follows that $\min_{z\in\{\lv,u\}}\{\dist_{\triangle_{q-1}'- D'}(z,x)\}\leq q - 1 + |D'| - 1 < q + |D|$. 
	If $z=\lv$, then we are done. 
	Thus let~$z=u$, where $u$ is the vertex incident to both $\lv$ and $\rv$ in~$\triangle_q$. 
	We know that~$\triangle_q- D$ is connected, and thus there exists an $u$-$\rv$~path in the right subfractal~$\triangle_{q-1}''- D''$. 
	By~\cref{lem:lengthofpathsbyone}, it follows that $\dist_{\triangle_{q-1}''- D''}(u,\rv)\leq |D''|+1$. 
	Recall that~$\{\lv,\rv\}\not\in D$. 
	In total, we get
	\begin{align*} 
		\dist_{\triangle_q- D}(x,\lv) 	& \leq \dist_{\triangle_{q-1}'- D'}(u,x) + \dist_{\triangle_{q-1}''- D''}(u,\rv) + 1  \\
															& \leq q-1+|D'|-1 + |D''|+1 + 1 = q + |D|.
	\end{align*} 
	In the cases, we obtain that~$\dist_{\triangle_q- D}(x,\lv) \le q + |D|$ and hence, $\dist_{\triangle_q- D}(x,\rv) \le q + |D| + 1$.
	In case that~$x$ is in the right subfractal~$\triangle_{q-1}''$, it follows by symmetry that~$\dist_{\triangle_q- D}(x,\rv) \le q + |D|$ and hence, $\dist_{\triangle_q- D}(x,\lv) \le q + |D| + 1$.
	
	\textbf{Case 2: $\{\lv,\rv\}\in D$.} First, we consider the step for statement (A). 
	Let $x$ be in the left subfractal~$\triangle_{q-1}'$. 
	Observe that $D'$ does not form an edge~cut in~$\triangle_{q-1}'$, since otherwise the graph is not connected. 
	Thus, $\triangle_{q-1}'- D'$ is connected, and by IH.(A) it follows that $\dist_{\triangle_{q-1}'- D'}(\lv,x)\leq q-1+|D'|+1 < q + |D|$. 
	
	Now, let~$x$ be in the right subfractal~$\triangle_{q-1}''$. 
	Again, $D''$~does not form an edge~cut in~$\triangle_{q-1}''$. 
	By IH.(A), $\dist_{\triangle_{q-1}''- D''}(u,x)\leq q-1+|D''|+1$. 
	Since $u$ and $\lv$ are connected in~$\triangle_{q-1}'- D'$, we can apply \cref{lem:lengthofpathsbyone} on $u$ and $\lv$. In total, with~$D=D'\cup D''\cup \{\lv,\rv\}$ we get: 
	\begin{align*} 
		\dist_{\triangle_q- D}(x,\lv) 	& \leq \dist_{\triangle_{q-1}'- D'}(u,\lv) + \dist_{\triangle_{q-1}''- D''}(u,x)  \\
															& \leq q-1+|D'|+1+|D''|+1 \leq q + |D|.
	\end{align*} 

	Next, we consider the step for statement (B). 
	Observe that the edge~cut formed by edges in~$D$ cannot form edge~cuts in $\triangle_{q-1}'$ and in $\triangle_{q-1}''$ at the same time since otherwise there are more than two connected components. 
	Let~$x$ be in the left subfractal and let edges in $D'$ do not form an edge~cut in~$\triangle_{q-1}'- D'$. 
	Then, by IH.(A), it follows that $\dist_{\triangle_{q-1}'- D'}(\lv,x)\leq q-1+|D'|+1 \le q + |D| - 1$. 
	Thus, let the edges in $D'$ form an edge~cut in~$\triangle_{q-1}'- D'$. 
	By IH.(B), either $\dist_{\triangle_{q-1}'- D'}(\lv,x)\leq q-1+|D'|-1 < q + |D| - 1$, or $\dist_{\triangle_{q-1}'- D'}(u,x)\leq q-1+|D'|-1$. 
	For the latter case, recall that the edges in $D''$ do not form a cut in $\triangle_{q-1}''$, that is, $\triangle_{q-1}''- D''$ is connected. 
	By~\cref{lem:lengthofpathsbyone}, it follows that $\dist_{\triangle_{q-1}''- D''}(u,\rv)\leq |D''|+1$. 
	In total, we get:
	\begin{align*} 
		\dist_{\triangle_q- D}(x,\rv) 	& \leq \dist_{\triangle_{q-1}'- D'}(x,u) + \dist_{\triangle_{q-1}''- D''}(u,\rv)  \\
															& \leq q-1+|D'|-1 + |D''|+1 < q+|D|-1.
	\end{align*} 
	The case where $x$ is in the right subfractal follows by symmetry. 
\end{proof}
\subsection{Directed Variants of~\Fractal{s}}\label{ssec:directedCase}

By definition, a~\fractal{} is an undirected graph. 
We now discuss how to turn it into a directed graph, more precisely, into a directed acyclic graph. 
We denote the directed variant of~$\triangle_q$ by~$\vec{\triangle}_q$. 
We obtain $\vec{\triangle}_q$ from $\triangle_q$ as follows:
Recall that each boundary forms a $\lv$-$\rv$~path. 
For each boundary, we direct the edges in the boundary from~$\lv$ to~$\rv$. 
By this, the obtained boundary forms a directed $\lv$-$\rv$~path. 
Observe that~$\lv$ has no incoming arcs, and the out-degree of~$\lv$ equals~$q+1$. 
Further observe that $\rv$ has no outgoing arcs, and the in-degree of $\rv$ equals $q+1$. 
Moreover, $\vec{\triangle}_q$ is acyclic, see \cref{fig:DirectedFractal} for an illustration.
\begin{figure}[t]
\centering
 \begin{tikzpicture}

\usetikzlibrary{decorations.pathmorphing}
\usetikzlibrary{decorations.pathreplacing}

\tikzstyle{tnode}=[fill, circle, scale=1/2, draw];
\tikzstyle{tnodespez}=[circle, scale=1/2, draw];
\tikzstyle{tnodesm}=[fill, circle, scale=1/3, draw];

\def\x{6};

\node (d1) at (0-\x,0)[tnodespez, label=270:{$\lv$}]{};
\node (d2) at (2-\x,0)[tnodespez, label=270:{$\rv$}]{};
\node (d3) at (1-\x,1)[tnode]{};
\draw[-stealth] (d1) to (d2); 
\draw[-stealth] (d1) to (d3);
\draw[-stealth] (d3) to (d2);

\node (d21) at (-0.5-\x,1.5)[tnode]{};
\node (d22) at (2+0.5-\x,1.5)[tnode]{};

\draw[-stealth] (d1) -- (d21);
\draw[-stealth] (d21) -- (d3);
\draw[-stealth] (d22) -- (d2);
\draw[-stealth] (d3) -- (d22);

\node (d31) at (-1.5-\x,0.75)[tnode]{};
\node (d32) at (0.25-\x,2.25)[tnode]{};
\node (d33) at (2-0.25-\x,2.25)[tnode]{};
\node (d34) at (2+1.5-\x,0.75)[tnode]{};

\draw[-stealth] (d1) -- (d31);
\draw[-stealth] (d31) -- (d21);
\draw[-stealth] (d32) -- (d3);
\draw[-stealth] (d21) -- (d32);
\draw[-stealth] (d3) -- (d33);
\draw[-stealth] (d33) -- (d22);
\draw[-stealth] (d34) -- (d2);
\draw[-stealth] (d22) -- (d34);

\end{tikzpicture}
\caption{The directed~\fractal{} $\vec{\triangle}_3$.}
\label{fig:DirectedFractal}
\end{figure}

Except for~\cref{lem:lengthofanyvertex}, all results from \cref{subsec:prop} can be transferred to~$\vec{\triangle}_q$. 
\cref{lem:MinCutOfSizeqPlusOne,cor:MinCutBoundaryEdges} hold since we still have the same degree on $\lv$ an $\rv$ and the boundaries still form disjoint (directed) $\lv$-$\rv$~paths. 
Furthermore, we still have the equivalent recursive definition with the adjustment that the edge between~$\lv$ and~$\rv$ becomes an arc from~$\lv$ to~$\rv$.
We define the dual structure of~$\vec{\triangle}_q$ as the dual structure of the underlying undirected variant~$\triangle_q$. 
By this, it is not hard to adapt~\cref{lem:path1to1cut,lem:DistToCompEnd}. 
For the latter result, and additionally for~\cref{lem:lengthofpathsbyone}, we make use of the fact that in the undirected case, we traverse the edges of the undirected~$\triangle_q$ in the same direction as they are directed in~$\vec{\triangle}_q$.

Regarding an equivalent of~\cref{lem:lengthofanyvertex} for the directed variant, with small effort one can modify the proof of~\cref{lem:lengthofanyvertex} to show the following.
\begin{lemma}\label[lemma]{lem:lengthofanyvertexdir}
	Let $D\subseteq E(\vec{\triangle}_q)$ be a subset of arcs of~$\vec{\triangle}_q$ and let~$x$ be an arbitrary vertex in~$V(\vec{\triangle}_q)$. 
	If~$x\in V(\vec{\triangle}_q)$ is reachable from~$\lv$ in~$\vec{\triangle}_q- D$, then $\dist_{\vec{\triangle}_q- D}(\lv,x)\leq q + |D| + 1$.
\end{lemma}

\begin{proof}
	We prove the statement with an induction on depth~$q$ of the~\fractal{}.
 
	The base case is~$q=0$.
	If $x=\lv$, the statement immediately holds.
	If $x=\rv$, observe that $D=\emptyset$, since $x$ is reachable from $\lv$.
	Thus, since $\rv$ has distance one to $\lv$, the statement follows. 
 
	As our induction hypothesis, we assume that the statement holds for $1,\ldots, q-1$. 
	We introduce some notation used for the induction step.
	Let~$\vec{\triangle}_q$, $q > 0$, be the directed \fractal with special vertices~$\lv$ and~$\rv$ and let~$u$ be the (unique) vertex in~$\vec{\triangle}_q$ that is adjacent to~$\lv$ and~$\rv$, that is, $u$ is on the boundary~$B_1$ of~$\vec{\triangle}_q$.
	Denote with~$\vec{\triangle}_{q-1}'$ ($\vec{\triangle}_{q-1}''$) the left (right) subfractal of~$\vec{\triangle}_q$ with special vertices~$\lv$ and~$u$ ($u$ and~$\rv$).
	Furthermore, let~$D'$ ($D''$) be the subset of arcs of~$D$ deleted in $\vec{\triangle}_{q-1}'$ ($\vec{\triangle}_{q-1}''$).

	Let $x\in V(\vec{\triangle})$ be an arbitrary vertex reachable from~$\lv$ in~$\vec{\triangle}_q- D$. 
	For the inductive step, we consider the two cases of the position of $x$ in $\vec{\triangle}_q$. 
 
	\textbf{Case 1: $x$ appears in $\vec{\triangle}_{q-1}'$.} 
	By induction hypothesis, it follows that
	\begin{align*} 
	 \dist_{\vec{\triangle}_q- D}(\lv,x) = \dist_{\vec{\triangle}_{q-1}'- D'}(\lv,x) \leq q - 1  + |D'| + 1 \leq q + |D|.
	\end{align*}

	\textbf{Case 2: $x$ appears in $\vec{\triangle}_{q-1}''$.} 
	Since $x$ is reachable from $\lv$ and $\rv$ has no outgoing arcs, it follows that $u$ is reachable from $\lv$ as well. 
	By the version of~\cref{lem:lengthofpathsbyone} for directed \fractal{s}, it follows that $\dist_{\vec{\triangle}_{q-1}'- D'}(\lv,u)\leq |D'|+1$. 
	Together with the induction hypothesis, it follows that
	\begin{align*} 
		\dist_{\vec{\triangle}_q- D}(\lv,x) &\leq \dist_{\vec{\triangle}_{q-1}'- D'}(\lv,u) + \dist_{\vec{\triangle}_{q-1}'- D'}(u,\rv) \\
		& \leq |D'|+1 + q - 1  + |D''| + 1 \leq q + |D| + 1.
	\end{align*} 
\end{proof}

Observe that the case that $x$ reaches $\rv$ is symmetric.
\subsection{Application Manual for~\Fractal{s}}\label{ssec:apps}

The aim of this subsection is to provide two general guidelines on how to use~\fractal{}s in cross-compositions. 
To this end, we introduce two general constructions--one for undirected graphs and one for directed graphs.
We start with the undirected case.

\begin{constr}\label{constr:constr1}
	Given $p=2^q$ instances $\I_1,\ldots,\I_p$ of an NP-hard graph problem, where each instance $\I_i$ has a unique source vertex~$s_i$ and a unique sink vertex~$t_i$. 
	\begin{compactenum}[(i)]
		\item Equip~$\triangle_q^c$ with some ``appropriate'' edge~cost $c\in \N$. 
		\item Let $v_0,\ldots,v_p$ be the vertices of the boundary~$B_q$, labeled by their distances to~$\lv$ in the $\lv$-$\rv$~path corresponding to~$B_q$ (observe that~$v_0=\lv$ and~$v_p=\rv$). 
		\item Incorporate each of the $p$~graphs of the input instances into $\triangle_q^c$ as follows: for each $i\in[p]$, merge~$s_i$ with vertex~$v_{i-1}$ in~$\triangle_q^c$ and merge~$t_i$ with~$v_i$ in~$\triangle_q^c$.
	\end{compactenum}
\end{constr}

\begin{figure}[t]
	\centering
	\begin{tikzpicture}

\tikzstyle{tnode}=[fill, circle, scale=1/2, draw];
\tikzstyle{tnodespez}=[circle, scale=1/2, draw];

\foreach \x in {1,2,...,8}{
\node (s1) at (1.5*\x-6.25,5)[tnode, label={$s_\x$}]{};
\node (t1) at (1.5*\x-6.25+1,5)[tnode, label={$t_\x$}]{};
\draw[-, dashed] (s1) to [out=45,in=135](t1);
\draw[-, dashed] (s1) to [out=-45,in=-135](t1);
}

\def\ya{1.25}

\draw[->] (-5.5+1.25,4.5) to [out=270, in=180](-1,0+\ya);
\draw[->] (-4+1.25,4.5) to [out=270, in=90](-1,1.5+\ya);
\draw[->] (-2.5+1.25,4.5) to [out=270, in=90](-0.25,2.25+\ya);
\draw[->] (-1+1.25,4.5) to [out=270, in=90](0.75,2+\ya);
\draw[->] (0.5+1.25,4.5) to [out=270, in=90](2-0.75,2+\ya);
\draw[->] (2+1.25,4.5) to [out=270, in=90](2+0.25,2.25+\ya);
\draw[->] (3.5+1.25,4.5) to [out=270, in=90](2+1,1.5+\ya);
\draw[->] (5+1.25,4.5) to [out=270, in=0](2+1,0+\ya);

\node (d1) at (0,0+\ya)[tnodespez, label=270:{$\lv$}]{};
\node (d2) at (2,0+\ya)[tnodespez, label=270:{$\rv$}]{};
\node (d3) at (1,1+\ya)[tnode]{};
\draw (d1) -- (d2) -- (d3) --(d1);

\node (d21) at (-0.5,1.5+\ya)[tnode]{};
\node (d22) at (2+0.5,1.5+\ya)[tnode]{};

\draw (d1) -- (d21) -- (d3);
\draw (d2) -- (d22) -- (d3);

\node (d31) at (-1.5,0.75+\ya)[tnode]{};
\node (d32) at (0.25,2.25+\ya)[tnode]{};
\node (d33) at (2-0.25,2.25+\ya)[tnode]{};
\node (d34) at (2+1.5,0.75+\ya)[tnode]{};

\draw (d1) -- (d31) -- (d21);
\draw (d3) -- (d32) -- (d21);
\draw (d3) -- (d33) -- (d22);
\draw (d2) -- (d34) -- (d22);

\draw[decorate, decoration={brace, amplitude=10pt}, thin] (6,-0.5+\ya)--(-4,-0.5+\ya);
\def\y{-3.5+\ya}

\node (d1) at (0,0+\y)[tnodespez, label=270:{$\lv$}]{};
\node (d2) at (2,0+\y)[tnodespez, label=270:{$\rv$}]{};
\node (d3) at (1,1+\y)[tnode]{};
\draw (d1) -- (d2) -- (d3) --(d1);

\node (d21) at (-0.5,1.5+\y)[tnode]{};
\node (d22) at (2+0.5,1.5+\y)[tnode]{};

\draw (d1) -- (d21) -- (d3);
\draw (d2) -- (d22) -- (d3);

\node (d31) at (-1.5,0.75+\y)[tnode]{};
\node (d32) at (0.25,2.25+\y)[tnode]{};
\node (d33) at (2-0.25,2.25+\y)[tnode]{};
\node (d34) at (2+1.5,0.75+\y)[tnode]{};

\draw (d1) -- (d31) -- (d21);
\draw (d3) -- (d32) -- (d21);
\draw (d3) -- (d33) -- (d22);
\draw (d2) -- (d34) -- (d22);

\draw[dashed] (d1) to [out=170, in=-60](d31);
\draw[dashed] (d1) to [out=190, in=-90](d31);

\draw[dashed] (d31) to [out=70, in=-160](d21);
\draw[dashed] (d31) to [out=90, in=-180](d21);

\draw[dashed] (d21) to [out=70, in=-160](d32);
\draw[dashed] (d21) to [out=90, in=-180](d32);

\draw[dashed] (d32) to [out=-10, in=110](d3);
\draw[dashed] (d32) to [out=10, in=100](d3);

\draw[dashed] (d3) to [out=80, in=180](d33);
\draw[dashed] (d3) to [out=60, in=-160](d33);

\draw[dashed] (d33) to [out=0, in=90](d22);
\draw[dashed] (d33) to [out=-20, in=110](d22);

\draw[dashed] (d22) to [out=0, in=90](d34);
\draw[dashed] (d22) to [out=-20, in=110](d34);

\draw[dashed] (d34) to [out=-90, in=0](d2);
\draw[dashed] (d34) to [out=-110, in=20](d2);

\end{tikzpicture}
	\caption{Illustration of~\cref{constr:constr1} with $p=2^3=8$. The vertices $s_1,\ldots,s_8$ indicate the source vertices in the eight input instances, and $t_1,\ldots,t_8$ indicate the sink vertices in the eight input instances. We use dashed lines to sketch the input graphs. 
	Below the curved brace, the resulting graph of the target instance is sketched.
	}
	\label{fig:crosscomp-spmve}
\end{figure}
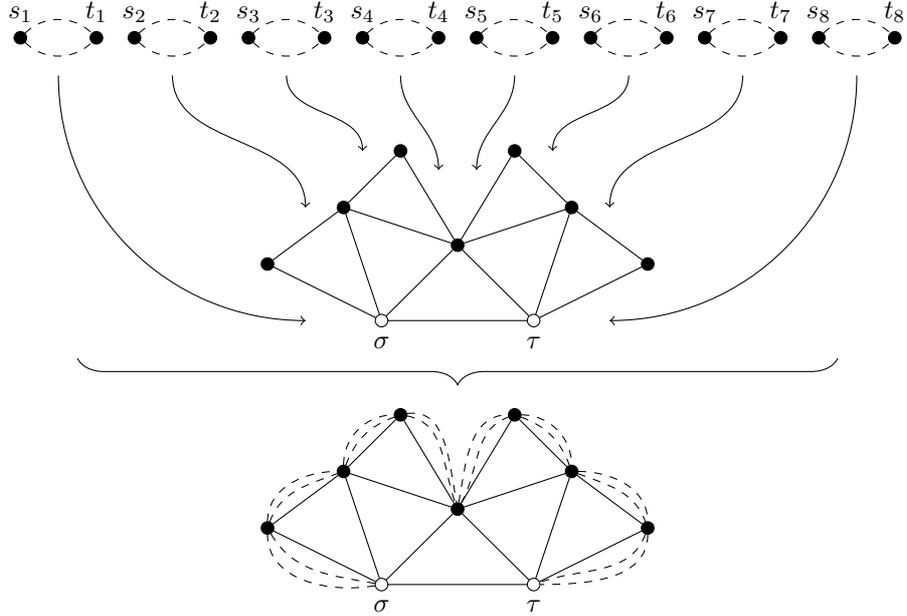

Refer to~\cref{fig:crosscomp-spmve} for an illustrative example of~\cref{constr:constr1}.
In \cref{constr:constr1}, the \fractal{} works as an instance selector by deleting edges corresponding to a minimum edge~cut, which, by \cref{lem:MinCutOfSizeqPlusOne}, is of size~$q+1$. 
Hence, each minimum edge~cut costs~$c \cdot (q+1)$.
The idea is that if we choose an appropriate value for~$c$ (larger than the budget in the instances~$\I_1,\ldots,\I_p$) and an appropriate budget in the composed instance (e.\,g.\ $c \cdot (q+1)$ plus the budget in the instances~$\I_1,\ldots,\I_p$), then we can only afford to delete at most~$q+1$ edges in~$\triangle_q^c$.
Furthermore, if the at most~$q+1$ edges chosen to be deleted do not form a minimum $\lv$-$\rv$ edge~cut in~$\triangle_q^c$, then, by \cref{lem:lengthofpathsbyone}, the shortest $\lv$-$\rv$~path has length at most~$q+2$.
Thus, by requiring in the composed instance that~$\lv$ and~$\rv$ have distance more than~$q+2$, we enforce that any solution for the composed instance contains a minimum $\lv$-$\rv$ edge~cut in~$\triangle_q^c$.
By~\cref{lem:path1to1cut}, each such minimum edge~cut corresponds to one root-leaf path in the dual structure~$T_q$ of~$\triangle_q^c$.
Observe that each leaf in the dual structure of~$\triangle_q^c$ one-to-one corresponds to an attached source instance. 
Hence, with an appropriate choice of~$c$, the budget in the composed instance, and the required distance between~$\lv$ and~$\rv$, the \fractal{} ensures that one instance is ``selected''.
We say that a minimum $\lv$-$\rv$ edge~cut in~$\triangle_q^c$ \emph{selects} an instance~$\I$ if the edge~cut corresponds to the root-leaf path with the leaf corresponding to instance~$\I$.

\begin{observation}\label{obs:intanceselection}
	Every minimum edge~cut~$C$ in~$\triangle_q^c$ selects exactly one instance~$\I$. 
	Conversely, every instance~$\I$ can be selected by exactly one minimum edge~cut.
\end{observation}

Moreover, the graph obtained from \cref{constr:constr1} has treewidth bounded in the maximum input instance size.

\begin{observation}\label{obs!constr1tw}
 Let $n_{\max}:=|V(G_i)|$, where $G_i$ is the graph in instance~$\I_i$, $i\in [p]$, from \cref{constr:constr1} and let $G$~be the obtained graph.
Then the treewidth of $G$ is $\tw(G)\leq 2+n_{\max}$.
\end{observation}

\begin{proof}
	By \cref{obs!outerplanar}, we know that the treewidth of \fractal{} is at most two. 
	Moreover, we know that the treewidth of the modified \fractal{} is at most two (cf.\ paragraph preceding \cref{subsec:prop}).
	Considering a tree decomposition of the modified \fractal{}, we replace each bag corresponding to an edge~$e$ of the outer boundary by the set containing all vertices of the instance appended on $e$. 
	Hence, we obtain a tree decomposition of~$G$ of width at most~$n_{\max}+2$.
\end{proof}

\begin{proposition}\label{prop!treewidth}
 Unless $\NP\subseteq \coNP/\poly$, any parameterized problem $P$ that admits an OR-cross-composition for some \NP-hard problem~$L$ by using \cref{constr:constr1} does not admit a polynomial kernel with respect to the parameter treewidth~$\tw$.
\end{proposition}

Using the same ideas as above and transferring them to the directed case yields the following construction with analogous properties.

\begin{constr}\label{constr:constr2}
 Given $p=2^q$ instances $\I_1,\ldots,\I_p$ of an NP-hard problem on directed acyclic graphs, where each instance~$\I_i$ has a unique source vertex~$s_i$ and a unique sink vertex~$t_i$. 
 \begin{compactenum}[(i)]
  \item Equip~$\vec{\triangle}_q^c$ with some ``appropriate'' edge~cost $c\in \N$, where $\lv$ is the vertex with no incoming arc.
  \item Let~$v_0,\ldots,v_p$ be the vertices of the boundary~$B_q$, labeled by their distances to~$\lv$ in the $\lv$-$\rv$~path corresponding to~$B_q$ (observe that~$v_0=\lv$ and~$v_p=\rv$).
  \item Incorporate each of the $p$~directed acyclic graphs of the input instances into~$\vec{\triangle}_q^c$ as follows: for each $i\in[p]$, merge~$s_i$ with vertex~$v_{i-1}$ in~$\vec{\triangle}_q^c$ and merge~$t_i$ with vertex~$v_i$ in~$\vec{\triangle}_q^c$.
 \end{compactenum}
\end{constr}

In the rest of the paper, we use~\cref{constr:constr1,constr:constr2} in OR-cross-com\-po\-si\-tions to rule out the existence of polynomial kernels. 
We baptize this approach \emph{fractalism}. 
In particular, we provide the source and the target problem, appropriate values for the edge~cost~$c$ and the budget in the composed instance, and the required distance between the special vertices~$\lv$ and~$\rv$. 
Observe that the directed graph obtained from~\cref{constr:constr2} is acyclic. 
Hence, by~\cref{constr:constr2} we can apply OR-cross-compositions for problems on directed acyclic graphs. 
We remark that there is a third construction where we drop the ``acyclicity'' requirement in~\cref{constr:constr2}. 
This yields a construction of a directed, possibly cyclic graph. 
In this sense, \cref{constr:constr2} is a special case of the third construction. 
\section{Applications to Length-Bounded Cut Problems}

In this section, we rule out the existence of polynomial kernels for several problems (and their variants) under the assumption that $\NP\not\subseteq \coNP/\poly$. 
To this end, we combine the framework of OR-cross-compositions with our fractalism technique as described in \cref{ssec:apps}. 

\subsection{\spmve}

Our first application of fractalism is the \spmveTsc problem~\cite{BaierEHKKPSS10}, also known as the problem of finding bounded edge undirected cuts \cite{GolovachT11}, or the \textsc{Shortest Path Most Vital Edges} problem \cite{malik1989k,BazganNN15}. %

\decprob{\spmve{} (\spmveAcr)}
	{An undirected graph $G=(V,E)$, with $s,t\in V$, and two integers~$k, \ell$.}
	{Is there a subset $F\subseteq E$ of cardinality at most $k$ such that $\dist_{G-F}(s,t)\geq \ell$?}

The problem is \NP-complete \cite{ItaiPS82} and fixed-parameter tractable with respect to~$(k,\ell)$~\cite{GolovachT11}. 
If $k$ is at least the size of any $s$-$t$~edge~cut, then the problem becomes polynomial-time solvable by simply computing a minimum $s$-$t$~edge cut. 
Thus, throughout this section, we assume that $k$ is smaller than the size of any minimum $s$-$t$~edge~cut. 
The generalized problem where each edge is equipped with positive length remains \NP-hard even on series-parallel and outerplanar graphs~\cite{BaierEHKKPSS10}. 
The directed variant with positive edge lengths remains \NP-hard on planar graphs where the source and the sink vertex are incident to the same face~\cite{PanS16}. 
Recently, Dvo\v{r}\'{a}k and Knop~\citet{DvorakK15} showed that the problem can be solved in polynomial time on graphs of bounded treewidth. 
Here, we answer an open question~\cite{GolovachT11} concerning the existence of a polynomial kernel with respect to the combined parameter $(k,\ell)$.\footnote{The question also appeared in the list of open problems of the FPT School 2014, 17-22 August 2014, B\k{e}dlewo, Poland, \url{http://fptschool.mimuw.edu.pl/opl.pdf}.}

\begin{theorem}\label{theo:spmve}
	Unless~$\NP\subseteq \coNP/\poly$, \spmveTsc{} parameterized by~$(k,\ell,\omega)$ does not admit a polynomial kernel, where~$\omega$ denotes the treewidth.
\end{theorem}
\begin{proof}
We OR-cross-compose $p=2^q$ instances of \spmveAcr{} into one instance of \spmveAcr{}$(k',\ell')$. 
An instance~$(G_i,s_i,t_i,k_i,\ell_i)$ of~\spmveAcr{} is called \emph{bad} if $\max\{k_i,\ell_i\}>|E(G_i)|$ or $\min\{k_i,\ell_i\}<0$. 
We define the polynomial equivalence relation~$\mathcal{R}$ on the instances of~\spmveAcr{} as follows: two instances $(G_i,s_i,t_i,k_i,\ell_i)$ and $(G_j,s_j,t_j,k_j,\ell_j)$ of \spmveAcr{} are $\mathcal{R}$-equivalent %
if and only if~$k_j=k_i$ and~$\ell_j=\ell_i$, or both are bad instances.
Clearly, the relation~$\mathcal{R}$ fulfills condition~(i) of an equivalence relation (see~\cref{sec:prelim}).
Observe that the number of equivalence classes of a finite set of instances is upper-bounded by the maximal size of a graph over the instances, hence condition~(ii) holds.
Thus, we consider $p$ $\mathcal{R}$-equivalent instances $\I_i:=(G_i,s_i,t_i,k,\ell)$, $i=1,\ldots,p$. 
We remark that we can assume that $\ell\geq 3$, since otherwise \spmveAcr{} is solvable in polynomial time by counting all edges connecting the source with the sink vertex. 
We OR-cross-compose into one instance $\I:=(G,s,t,k',\ell')$ of \spmveAcr{}$(k',\ell')$ with $k'=k^2\cdot(\log(p)+1)+k$ and $\ell'=\ell+\log(p)$ as follows. 

\emph{Construction}: Apply~\cref{constr:constr1} with edge~cost $c=k^2$. 
In addition, set $s:=\lv$ and $t:=\rv$. 
Let~$G$~denote the obtained graph. 
By \cref{obs!constr1tw}, the treewidth $\tw(G)$ of~$G$ is at most $2+\max_{i\in[p]}|V(G_i)|$.

\emph{Correctness}: We show that $\I$~is a \yes-instance if and only if there exists an~$i\in[p]$ such that $\I_i$~is a \yes-instance. 

``$\Leftarrow$'': Let $i\in[p]$ be such that $\I_i$ is \yes. 
Following \cref{obs:intanceselection} in \cref{ssec:apps}, let~$C$~be the minimum $s$-$t$~cut in $\triangle_q^c$ that selects instance~$\I_i$. 
Recall that $C$ is of size~$q+1$ and that the edge~cost equals~$k^2$. 
Thus, the minimum $s$-$t$~cut~$C$ has cost~$(q+1)\cdot k^2=(\log(p)+1)\cdot k^2$.

Note that after deleting the edges in~$C$, the vertices~$s$ and $t$ are only connected via paths through the incorporated graph~$G_i$. 
Since $\I_i$ is \yes, we can delete $k$~edges (equal to the remaining budget) such that the distance of~$s_i$ and~$t_i$ in~$G_i$ is at least~$\ell$. 
Together with \cref{lem:DistToCompEnd} in \cref{subsec:prop}, such an additional edge deletion increases the length of any shortest $s$-$t$~path in~$G$ to at least $\ell+\log(p)=\ell'$. 
Hence, $\I$ is a \yes-instance. 

``$\Rightarrow$'': Suppose that one can delete at most $k'$ edges in $G$ such that each $s$-$t$~path is of length at least $\ell'$. 
Since the budget allows $\log(p)+1$ edge-deletions in $\triangle_q^c$, by \cref{lem:lengthofpathsbyone} in \cref{subsec:prop}, if we do not cut~$s$ and~$t$ in~$\triangle_q^c$, then there is an $s$-$t$~path of length~$\log(p)+2$. 
Since~$\ell\geq 3$, such an edge deletion does not yield a solution. 
Thus, in every solution of~$\I$, a subset of the deleted edges forms a minimum $s$-$t$~edge~cut in $\triangle_q^c$ and thus, by \cref{obs:intanceselection}, selects an input instance. 

Consider an arbitrary solution to~$\I$, that is, an edge subset of~$E(G)$ of cardinality at most~$k'$ whose deletion increases the shortest $s$-$t$ path to at least~$\ell'$.
Let~$\I_i$, $i\in[p]$,~be the selected instance. 
Note that any shortest $s$-$t$~path contains edges in the selected instance~$\I_i$. 
By~\cref{lem:DistToCompEnd}, we know that the length of the shortest $s$-$s_i$~path and the length of the shortest $t_i$-$t$~path sum up to exactly $\log(p)$. 
It follows that the remaining budget of~$k$ edge deletions is spent in~$G_i$ in such a way that there is no path from $s_i$ to $t_i$ of length smaller than~$\ell$ in~$G_i$. 
Hence, $\I_i$~is a \yes-instance.
\end{proof}
Golovach and Thilikos~\citet{GolovachT11} showed that \spmveAcr{} on directed acyclic graphs is \NP-complete. 
Using \cref{constr:constr2} instead of \cref{constr:constr1} with \spmveAcr{} on directed acyclic graphs as source and target problem, the same argument as in the proof of~\cref{theo:spmve} yields the following.

\begin{theorem}\label{thm:spmveondags}
Unless~$\NP\subseteq \coNP/\poly$, \spmveTsc{} on directed acyclic graphs parameterized by $(k,\ell)$ does not admit a polynomial kernel.
\end{theorem}

In the following, we consider \spmveAcr{} on planar graphs. 
To the best of our knowledge, it was not shown before whether \spmveAcr{} remains \NP-hard on planar graphs. 
This is what we state next. 

\begin{theorem}\label{thm:planarspmveisnphard}
	\spmveTsc{} is \NP-hard even on planar undirected graphs as well as on planar directed acyclic graphs, where for both problems $s$ and $t$ are incident to the outer face.
\end{theorem}
To prove the theorem, we need the following definitions of plane embeddings of graphs.

\begin{definition}
 A \emph{page embedding} of a graph~$G$ is a plane embedding of~$G$ where all vertices lie on the real line and every edge lies in the upper half $\R\times \R^+$.
\end{definition}

\begin{definition}\label{def!bookembed}
 A graph $G=(V,E)$ is \emph{$k$-page book embeddable} if there is a partition~$E_1,\ldots,E_k$ of the edge set~$E$ such that $G_i:=(V,E_i)$ is page embeddable for all $i\in[k]$.
\end{definition}

Intuitively, a book embedding of a graph is a drawing of graph where all vertices are drawn along the spine of the book, and the edges are draw crossing-free on each page of the book.

\begin{proof}[Proof of \cref{thm:planarspmveisnphard}]
  Our proof follows the same strategy as the proof due to Schieber et al.~\citet{bar1998complexity} for \spmveAcr{} on general graphs, where Schieber et al.~\citet{bar1998complexity} reduce \textsc{Vertex Cover} to \spmveAcr{}.
  We reduce from \textsc{3-Planar Vertex Cover}, that is, \textsc{Vertex Cover} on planar graphs with maximum degree three, which remains \NP-complete~\cite{Mohar01}. 
  Heath~\citet{heath1985algorithms} proved that any planar graph of maximum vertex degree three allows a two-page embedding (cf.~\cref{def!bookembed}). 
  Moreover, Heath~\citeauthor{heath1985algorithms} showed that such an embedding can be computed in linear time in the number of vertices of the input graph. 
  Recently, Bekos et al.~\citet{BekosGR14} proved that any planar graph of maximum degree four allows a two-page embedding. 
  We mainly copy the proof due to Schieber et al.~\citet{bar1998complexity} and, on the way, perform small changes on the gadgets and target parameters. 
  We describe this in the following.
  \begin{figure}[t]
  \centering
  \begin{tikzpicture}

  \tikzstyle{tnode}=[circle,scale=1/2,draw];

  \def\xl{0.78};
  \def\xm{0.55};
  \def\y{1.25};

  \node (s) at (0,0)[tnode, label=135:$s_i$]{};
  \node (t) at (7*\xm, 0)[tnode, label=-45:$t_i$]{};

  \node (u2) at (2*\xl,\y)[tnode, label=-90:{\scriptsize{$x_i^u$}}]{};
  \draw[-] (s) to node [above, sloped]{$k-1$}(u2);
  \draw[-, dotted, very thick] (s) to (u2);
  \node (u3) at (3*\xl,\y)[tnode, label=-90:{\scriptsize{$y_i^u$}}]{};
  \draw (u2) -- (u3);
  \draw (t) -- node [above, sloped]{$k-1$}(u3);
  \draw[dotted, very thick] (t) -- (u3);

  \node (u2) at (2*\xl,-\y)[tnode, label=90:{\scriptsize{$x_i^\ell$}}]{};
  \draw (s) -- node [below,sloped]{$k-1$}(u2);
  \draw[dotted, very thick] (s) -- (u2);
  \node (u3l) at (3*\xl,-\y)[tnode, label=90:{\scriptsize{$y_i^\ell$}}]{};
  \draw (u2) -- (u3l);
  \draw (t) -- node [below, sloped]{$k-1$}(u3l);
  \draw[dotted, very thick] (t) -- (u3l);

  \draw (t) -- node [midway, above]{$2k$} (s);
  \draw[dotted, very thick] (t) -- (s);

  \def\xs{5.5};

  \node (ld) at (\xs-\xs/7,0)[]{$\ldots$};

  \node (s) at (0+\xs,0)[tnode, label=135:$s_j$]{};
  \node (t) at (7*\xm+\xs, 0)[tnode, label=-45:$t_j$]{};

  \node (u2) at (2*\xl+\xs,\y)[tnode, label=-90:{\scriptsize{$x_j^u$}}]{};
  \draw[-] (s) to node [above, sloped]{$k-1$}(u2);
  \draw[-, dotted, very thick] (s) to (u2);
  \node (u3) at (3*\xl+\xs,\y)[tnode, label=-90:{\scriptsize{$y_j^u$}}]{};
  \draw (u2) -- (u3);
  \draw (t) -- node [above, sloped]{$k-1$}(u3);
  \draw[dotted, very thick] (t) -- (u3);

  \node (u2r) at (2*\xl+\xs,-\y)[tnode, label=90:{\scriptsize{$x_j^\ell$}}]{};
  \draw (s) -- node [below,sloped]{$k-1$}(u2r);
  \draw[dotted, very thick] (s) -- (u2r);
  \node (u3) at (3*\xl+\xs,-\y)[tnode, label=90:{\scriptsize{$y_j^\ell$}}]{};
  \draw (u2r) -- (u3);
  \draw (t) -- node [below, sloped]{$k-1$}(u3);
  \draw[dotted, very thick] (t) -- (u3);

  \draw (t) -- node [midway, above]{$2k$} (s);
  \draw[dotted, very thick] (t) -- (s);

  \draw[-] (u3l) to [out=-90,in=-90] node [midway,below]{$(2k-1)\cdot (j-i)-2$} (u2r); 
  \draw[dotted, very thick] (u3l) to [out=-90,in=-90](u2r);

  \node (ld) at (-0.75,0)[]{$\ldots$};
  \node (ld) at (\xs+7*\xm+0.75,0)[]{$\ldots$};

  \end{tikzpicture}
  \caption{Illustration of the gadgets in the proof of~\cref{thm:planarspmveisnphard}. Here, exemplified for two vertices $i,j\in V$ with $\{i,j\}\in E$, and the edge is embedded on the second (lower) page in the two-page embedding of the input graph~$G=(V,E)$.}
  \label{fig:gadgetsplanarspmvenphard}
  \end{figure}
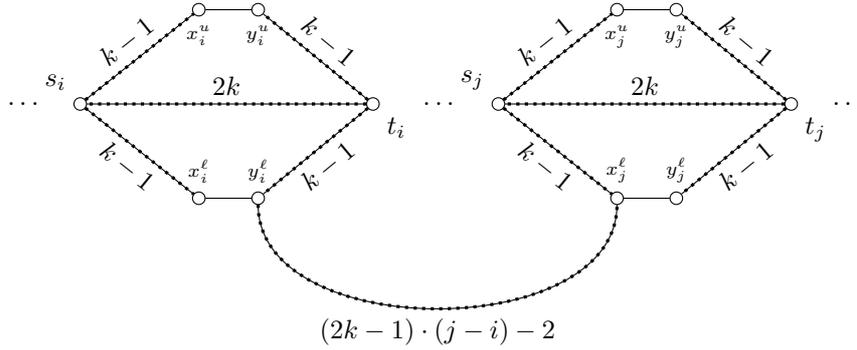

  Let~$\I=(G,k)$ be an instance of \textsc{3-Planar Vertex Cover}. 
  Since we can assume to have a two-page embedding, the vertices are drawn along the real line and connected by non-crossing edges lying in the lower and upper half. 
  Further, we assume that the vertices are labeled from~$1$ to~$n$, in the order along the real line.  
  We replace each vertex $i$ by a gadget~$\mathbf{i}$ as follows.  
  The gadget~$\mathbf{i}$ consists of two $P_{2k}$s, where $P_{2k}$~denotes a simple path with~$2k$ vertices, and one $P_{2k+1}$, all three merged together at their endpoints. 
  We denote the left and right (merged) endpoint of gadget~$\mathbf{i}$ by~$s_i$ and~$t_i$, respectively. 
  One $P_{2k}$ belongs to the upper half, the other to the lower half. 
  The $P_{2k+1}$ lies along the real line. 
  The two middle vertices of each of the two $P_{2k}$ we denote by $x_i^u, x_i^\ell, y_i^u, y_i^\ell$, where $x$ is left of $y$, and $u$ and $\ell$ stand for ``upper'' and ``lower''.
  We merge $t_i$ with $s_{i+1}$ for all $i\in[n-1]$. 
  We set $s:=s_1$ and $t:=t_n$. 
  Moreover, if two vertices $i<j$ are connected by an edge lying in the upper half, then we connect the vertex~$y_i^u$ with~$x_j^u$ via a path of length $(2k-1)(j-i)-2$ (analogously for edges in the lower half). 
  Refer to~\cref{fig:gadgetsplanarspmvenphard} for an illustration of the construction. 
  We denote by~$G'$ the obtained graph. 
  Observe that~$G'$ remains planar.
  Except for the edges~$\{x_i^u, y_i^u\}, \{x_i^\ell,y_i^\ell\}$, $i\in[n]$, there are no edges that are allowed to be deleted (see Schieber et al.~\citet{bar1998complexity}).

  We set $k':=2k$ and $\ell'=k\cdot (2k) + (n-k)\cdot (2k-1)$. 
  Let $\I':=(G',s,t,k',\ell')$ be the resulting instance of \textsc{Planar-\spmveAcr{}}$(k',\ell')$, that is \spmveAcr{$(k',\ell')$} on planar graphs. 
  We show that $\I$ is a \yes-instance if and only if $\I'$ is a \yes-instance.

  ``$\Rightarrow$'': 
  Suppose that $G$ admits a vertex cover of size at most~$k$. 
  Let $C\subseteq V(G)$ be such a vertex cover of size~$k$. 
  We claim that deleting the edges in the edge set $X:=\{\{x_i^u, y_i^u\}, \{x_i^\ell,y_i^\ell\}\mid i\in C\}$ forms a solution to~$\I'$. 

  We observe that any $s$-$t$~path in~$G'-X$ using only edges in the gadgets is of length at least~$\ell'$. 
  To see this, consider a gadget~$\mathbf{i}$ with $i\in C$. 
  Then the edges $\{x_i^u, y_i^u\}, \{x_i^\ell,y_i^\ell\}\in X$, and hence the only $s_i$-$t_i$~path using only edges in the gadget~$\mathbf{i}$ is of length~$2k$ (that is the~$P_{2k+1}$ used in the construction).
  If no edge in a gadget~$\mathbf{j}$ is deleted, then any shortest $s_j$-$t_j$~path using only edges in the gadget~$\mathbf{j}$ is of length~$2k-1$ (those correspond to the~$P_{2k}$s used in the construction).
  Since $|C|=k$, any $s$-$t$~path in~$G'-X$ using only edges in the gadgets is of length at least~$k\cdot(2k)+(n-k)\cdot (2k-1)=\ell'$.
  
  We have to show that there is no shorter $s$-$t$~path in~$G'-X$ than any path using only edges in the gadgets. 
  To this end, let $i,j\in V(G)$, $i<j$, be two adjacent vertices in $G$, that is, with $\{i,j\}\in E(G)$. 
  Since $C$ is a vertex cover, it follows that either $i\in C$ or $j\in C$. 
  Let~$i\in C$ and $j\not\in C$ (the case with~$j\in C$ and $i\not\in C$ is symmetric). 
  We consider the shortest path from $s_i$ to $t_j$ not going backwards, that is, not appearing in any gadget~$z$ with $z<i$ or $z>j$, and using the path connecting the gadgets of~$i$ and $j$. 
  Let the path connecting the gadgets of~$i$ and $j$ be a lower path, that is, the vertices~$y_i^\ell$ and~$x_j^\ell$ are connected by the path. 
  Since the edges $\{x_i^\ell, y_i^\ell\}$ and~$\{x_i^u, y_i^u\}$ are deleted, the shortest path from $s_i$ to $y_i^\ell$ is of length $2k+(k-1)$. 
  Then we take the path of length $(2k-1)(j-i)-2$ to get to the gadget of~$j$. 
  Finally, we take the path from~$x_j^\ell$ via edge~$\{x_j^\ell, y_j^\ell\}$ to $t_i$ of length $(k-1)+1=k$. 
  In total, the path is of length $4k+(2k-1)(j-i)-3$, and it is the shortest of its kind. 

  We compare this to the shortest path from $s_i$ to $t_j$ using only edges in the gadgets. 
  The length of such a path is at most $2k(k)+(2k-1)(j-i-k) +(2k-1) = 2k(j-i) - (j-i-k) + (2k-1)$ if $j-i\geq k$, and at most~$2k(j-i)+(2k-1)$ otherwise. 
  Comparing the two lengths, we obtain for~$j-i \ge k$
  \begin{align*}
	  4k+(2k-1)(j-i)-3 - (2k(j-i)-(j-i-k)+(2k-1)) &= k - 2,
  \end{align*}
  and for~$j - i < k$
  \begin{align*}
	  4k+(2k-1)(j-i)-3 - (2k(j-i)+(2k-1)) &= 2k - (j-i) - 2 > k-2.
  \end{align*}

  It follows that there is a path using only edges in the gadgets that is shorter than the shortest paths using at least one edge not appearing in the gadgets. 
  Finally note that if both $i,j\in C$, then the difference of the path lengths is even bigger. 
  Observe that using a path connecting gadget~$\mathbf{i}$ with~$\mathbf{j+1}$ (or $\mathbf{i-1}$ with~$\mathbf{j}$) to get from $s_i$ to $t_j$ is longer by at least $k-1$ (or at least $k-3$), following from an analogous argumentation as above. 
  Hence, the shortest path connecting $s$ with $t$ passes through the gadgets and is of length at least $\ell'$.

  ``$\Leftarrow$'': Suppose that $G'$ allows $k'=2k$ edge deletions such that any shortest $s$-$t$~path is of length at least $\ell'$. 
  Our first observation is that in any solution to~$\I$, either none or exactly two edges are deleted in any gadget. 
  Suppose that there is an gadget with only one edge deleted. 
  Then a shortest path through this gadget is of length $2k-1$. 
  Since $2k$ is the maximum increase of the passing length through a gadget, we get $(2k)(k-2)+(2k-1)(n-k+2)<(2k)\cdot k+(2k-1)(n-k) = \ell'$. 
  Hence, in any gadget, either exactly two or no edge is deleted. 
  Let $C\subseteq V(G)$ be the set of vertices such that both edges are deleted in the corresponding gadgets. 
  We claim that~$C$ is a vertex cover of size~$k$ in~$G$.

  Suppose that there are two gadgets~$\mathbf{i}$ and~$\mathbf{j}$ not containing any deleted edge, that is, $\{i,j\}\cap C=\emptyset$, but $\{i,j\}\in E(G)$. 
  Then the shortest $s_i$-$t_j$~path using the path corresponding to edge~$\{i,j\}\in E(G)$ is of length~$2k+ (2k-1)(i-j)-2$. 
  The shortest $s_i$-$t_j$~path through the gadgets only is of length at least~$2k-1+(2k-1)(i-j)$. 
  Thus, the path using the path connecting the gadgets~$\mathbf{i}$ and~$\mathbf{j}$ is too short by exactly one, and hence, the shortest $s$-$t$~path is of length smaller than~$\ell'$. 
  This contradicts the fact that~$\{\{x_i^u, y_i^u\}, \{x_i^\ell,y_i^\ell\}\mid i\in C\}$ forms a solution to~$\I'$. 
  It follows that for each edge $\{i,j\}\in E(G)$ we have $|C\cap \{i,j\}|>0$. 
  This is exactly the property of a vertex cover, and thus, $C$~is a vertex cover in $G$ of size~$k$.

  We have shown that the problem is \NP-hard on planar, undirected graphs. 
  Observe that we can direct all edges from ``left to right''. 
  The planarity still holds, and we obtain a directed acyclic graph. 
  Since we have shown in the proof that ``going backwards'' is never optimal, the proof can be easily adapted. 
  Thus, the problem remains \NP-hard on planar directed acyclic graphs.
\end{proof}

Due to~\cref{thm:planarspmveisnphard}, we can use \spmveAcr{} on planar undirected graphs as well as on planar directed acyclic graphs, where in both cases the source and sink vertices are incident to the outer face, as source problem for OR-cross-compositions. 
The property that the source and the sink vertices are allowed to be incident with the same face in the input graph allows us to use \cref{constr:constr1,constr:constr2} with a target problem on planar graphs. 
Hence, together with the same argumentation as in the proof of~\cref{theo:spmve}, we obtain the following.

\begin{theorem}\label{thm:spmvenopolykernelforplanardag}
Unless~$\NP\subseteq \coNP/\poly$, \spmveTsc{} on planar undirected graphs parameterized by $(k,\ell,\omega)$ as well as on planar directed acyclic graphs parameterized by $(k,\ell)$ do not admit a polynomial kernel, where~$\omega$ denotes the treewidth.
\end{theorem}
\subsection{\dmve{}}

Our second fractalism application concerns a problem introduced by Schoone et al.~\citet{SchooneBL87}.

\decprob{\dmve{} (\dmveAcr{})}
	{A connected, undirected graph $G=(V,E)$, two integers $k, \ell$.}
	{Is there a subset $F\subseteq E$ of cardinality at most $k$ such that $G-F$ is connected and $\diam(G-F)\geq \ell$?}
The problem was shown to be NP-complete, also on directed graphs~\cite{SchooneBL87}.
A simple search tree algorithm yields fixed-parameter tractability with respect to~$(k,\ell)$:

\begin{theorem}\label{thm:dmvekellfpt}
	\dmveTsc{} can be solved in $O((\ell-1)^k n^2 (n+m))$ time.
\end{theorem}
\begin{proof}
	We give a search tree algorithm branching over the possible edge deletions to prove that \dmveAcr{}($k,\ell$) is fixed-parameter tractable. 
	The underlying crucial observation is that if some instance~$(G,k,\ell)$ of \dmveAcr{}($k,\ell$) is a \yes-instance, then there exists at least one pair of vertices~$v,w\in V$ in the graph~$G-X$ such that $\dist_{G-X}(v,w)\geq \ell$, where $X$ is a solution to~$(G,k,\ell)$. 
	Hence, we want to check whether we can increase by at most~$k$ edge deletions the length of any shortest path between the chosen pair up to at least~$\ell$, where we delete an edge only if its deletion leaves the graph connected. 
	
	To this end, for each pair, we apply the branching algorithm provided by Golovach and Thilikos~\citet{GolovachT11}: Find a shortest path and if its length is at most~$\ell - 1$, then branch in all cases of deleting an edge on this path and decrease~$k$ by one.
	In each branch, we need to check whether the graph is still connected. 
	This can be done in~$O(n+m)$ time with a simple depth/breadth first search.
	Hence, in total we obtain a branching algorithm running in~$O(n^2\cdot (\ell-1)^k(n+m))$ time. 
	Thus, \dmveAcr{}($k,\ell$) is fixed-parameter tractable.
\end{proof}

Complementing the fixed-parameter tractability of~\dmveAcr{}($k,\ell$), we show the following.
 
\begin{theorem}\label{thm:NoPolyKernel-D-MVE}
Unless~$\NP\subseteq \coNP/\poly$, \dmveTsc{} parameterized by $(k,\ell,\omega)$ does not admit a polynomial kernel, where~$\omega$ denotes the treewidth.
\end{theorem}
\begin{proof} 
We OR-cross-compose $p=2^q$ instances of \spmveTsc{} (\spmveAcr) on connected graphs into one instance of \dmveAcr{}$(k,\ell)$ as follows.
Apply~\cref{constr:constr1} with $p=2^q$ instances $\I_i:=(G_i,s_i,t_i,k,\ell)$, $i=1,\ldots,p$, of the input problem \spmveAcr{} on connected graphs, target problem~\dmveAcr{}$(k,\ell)$, and edge~cost~$c=k^2$. 
The equivalence relation on the input instances is defined as in the proof of~\cref{theo:spmve}. 
Let $n_{\max} :=\max_{i\in[p]} |V(G_i)|$. 
In addition, attach to~$\lv$ as well as on~$\rv$ a path of length~$L:=n_{\max}\cdot (2\log(p)+3)+1$ each. 
Denote the endpoint of the path attached to $\lv$ by $\lv'$ (where $\lv'\neq \lv$), and let $\rv'$ be defined analogously. 
Let~$G$ denote the obtained graph.
Note that the appended paths to the \fractal{} do not increase the treewidth $\tw(G)$ of $G$, and hence by \cref{obs!constr1tw}, it holds that $\tw(G)\leq 2+n_{\max}$.

Refer to~\cref{fig:crosscomp-dmve} for an exemplified illustration of the described construction. 
Let $\I:=(G,k',\ell')$~be the target instance of~\dmveAcr{}$(k',\ell')$ with~$k'=k^2\cdot(\log(p)+1)+k$ and~$\ell'=2\cdot L+\log(p)+\ell$.
 \begin{figure}[t]
\centering
 \begin{tikzpicture}

\tikzstyle{tnode}=[fill, circle, scale=1/2, draw];
\tikzstyle{tnodespez}=[circle, scale=1/2, draw];

\def\y{0}

\node (d1) at (0,0+\y)[tnodespez, label=225:{$\lv$}]{};
\node (d2) at (2,0+\y)[tnodespez, label=-45:{$\rv$}]{};
\node (d3) at (1,1+\y)[tnode]{};
\draw (d1) -- (d2) -- (d3) --(d1);

\node (d1b) at (0,-2)[tnode, label=225:{$\lv'$}]{};
\draw (d1) -- (d1b);
\foreach \x in {0.4,0.8,...,2} {
\node (temp) at(0,-\x)[tnode]{};
}

\draw[decorate, decoration={brace, amplitude=10pt}, thin] (0,0)--(0,-2);
\draw[decorate, decoration={brace, amplitude=10pt}, thin] (2,-2)--(2,0);
\node (txt) at (1,-1)[]{$L$};

\node (d2b) at (2,-2)[tnode, label=-45:{$\rv'$}]{};
\draw (d2) -- (d2b);
\foreach \x in {0.4,0.8,...,2} {
\node (temp) at(2,-\x)[tnode]{};
}

\node (d21) at (-0.5,1.5+\y)[tnode]{};
\node (d22) at (2+0.5,1.5+\y)[tnode]{};

\draw (d1) -- (d21) -- (d3);
\draw (d2) -- (d22) -- (d3);

\node (d31) at (-1.5,0.75+\y)[tnode]{};
\node (d32) at (0.25,2.25+\y)[tnode]{};
\node (d33) at (2-0.25,2.25+\y)[tnode]{};
\node (d34) at (2+1.5,0.75+\y)[tnode]{};

\draw (d1) -- (d31) -- (d21);
\draw (d3) -- (d32) -- (d21);
\draw (d3) -- (d33) -- (d22);
\draw (d2) -- (d34) -- (d22);

\draw[dashed] (d1) to [out=170, in=-60](d31);
\draw[dashed] (d1) to [out=190, in=-90](d31);

\draw[dashed] (d31) to [out=70, in=-160](d21);
\draw[dashed] (d31) to [out=90, in=-180](d21);

\draw[dashed] (d21) to [out=70, in=-160](d32);
\draw[dashed] (d21) to [out=90, in=-180](d32);

\draw[dashed] (d32) to [out=-10, in=110](d3);
\draw[dashed] (d32) to [out=10, in=100](d3);

\draw[dashed] (d3) to [out=80, in=180](d33);
\draw[dashed] (d3) to [out=60, in=-160](d33);

\draw[dashed] (d33) to [out=0, in=90](d22);
\draw[dashed] (d33) to [out=-20, in=110](d22);

\draw[dashed] (d22) to [out=0, in=90](d34);
\draw[dashed] (d22) to [out=-20, in=110](d34);

\draw[dashed] (d34) to [out=-90, in=0](d2);
\draw[dashed] (d34) to [out=-110, in=20](d2);

\end{tikzpicture}
\caption{Cross-composition for \dmveTsc{}($k,\ell$) with $p=8=2^3$, and $L=9\cdot n_{\max}+1$. Dashed lines sketch the boundaries of the graphs in the $p$ input instances.}
\label{fig:crosscomp-dmve}
\end{figure}
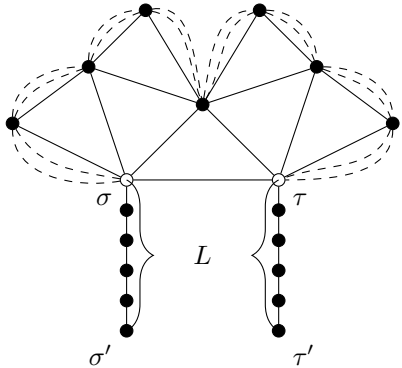

\emph{Correctness:} We show that $\I$ is a \yes-instance for \dmveAcr{}$(k',\ell')$ if and only if there exists an~$i\in[p]$ such that $\I_i$ is a \yes-instance for \spmveAcr{} on connected graphs.

``$\Leftarrow$'': Let $\I_i$, $i\in[p]$, be a \yes-instance for \spmveAcr{} on connected graphs. 
Following \cref{obs:intanceselection}, we delete all edges in the minimum cut in $\triangle_q^c$ that selects instance~$\I_i$. 
Then, we delete edges corresponding to a solution for~$\I_i$ without disconnecting the graph~$G$ (observe that we can always find such a solution). 
Let $X\subseteq E(G)$ be the set of deleted edges. 
The distance of~$\lv$ and~$\rv$ in~$G-X$ is at least~$\log(p)+\ell$, and thus, the distance of~$\lv'$ and~$\rv'$ is at least~$2\cdot L+\log(p)+\ell=\ell'$. 
Hence, the diameter is at least $\ell'$ after $k'$~edge deletions that leave the graph connected. 
It follows that $\I$~is a \yes-instance.
 
``$\Rightarrow$'': Conversely, suppose that~$\I$~allows $k'$~edge deletions such that the remaining graph is connected and has diameter at least~$\ell'$. 
Let~$X\subseteq E(G)$ be a solution. 
First observe that $G-X$ is connected. 
Consider the instances appended to the \fractal{} as the artificial $q+1$st boundary of a $(q+1)$-\fractal{}, where an edge in this boundary has length $n_{\max}$. 
Thus, we can apply~\cref{lem:lengthofanyvertex}(A) to this artificial $(q+1)$-\fractal{}. 
Recall that our budget only allows $\log(p)+1$ edge deletions (of cost $k^2$) in~$\triangle_q^c$. 
Hence we get that the distance to $\lv$ (and by symmetry to $\rv$) of every vertex contained either in $\triangle_q^c$ or in any appended instance is at most $n_{\max}\cdot (\log(p) + \log(p) + 3) = L-1$. 
It follows that $\dist_{G-X}(x,\lv)\leq \dist_{G-X}(\lv,\lv')$ and $\dist_{G-X}(x,\rv)\leq \dist_{G-X}(\rv,\rv')$ for all $x\in V(G)$. 
Moreover, for all $x,y\in V(G)$ we have:
\begin{align*}
 \dist_{G-X}(x,y) &\leq \dist_{G-X}(x,\lv) + \dist_{G-X}(\lv,\rv) + \dist_{G-X}(\rv,y) \\
 &\leq \dist_{G-X}(\lv',\lv) + \dist_{G-X}(\lv,\rv) + \dist_{G-X}(\rv,\rv') \\
 &= \dist_{G-X}(\lv',\rv').
\end{align*}

Hence, $\lv',\rv'$ is the pair of vertices with the largest distance in~$G-X$ and, thus, $\dist_{G-X}(\lv',\rv') \ge \ell'$.
Observe that $\dist_{G-X}(\lv',\rv')\geq \ell'$ if and only if $\dist_{G-X}(\lv,\rv)\geq \log(p)+\ell$ since every shortest $\lv'$-$\rv'$~path contains both~$\lv$ and~$\rv$. 
Following the argumentation in the correctness proof of~\cref{theo:spmve}, it follows that there is an instance $\I_i$, $i\in [p]$, that is a \yes-instance for \spmveAcr{} on connected graphs.
\end{proof}

In their \NP-hardness-proof for~\dmveAcr{}, Schoone et al.~\citet{SchooneBL87} reduce from \textsc{Hamiltonian Path (HP)} to~\dmveAcr{}. 
The reduction does not modify the graph, that is, the input graph for HP remains the same for the \dmveAcr{}~instance. 
Since HP remains \NP-hard on planar graphs~\cite{GareyJS76}, the reduction of~Schoone et al.\ implies that \dmveAcr{} is \NP-hard even on planar graphs. 
Due to~\cref{thm:planarspmveisnphard}, using \cref{constr:constr1} with \spmveAcr{} on planar graphs as source problem, and \dmveAcr{}$(k,\ell)$ on planar graphs as target problem, we obtain the following.

\begin{theorem}\label{cor:dmvenopkplanar}
Unless~$\NP\subseteq \coNP/\poly$, \dmveTsc{} on planar graphs parameterized by $(k,\ell,\omega)$ does not admit a polynomial kernel, where~$\omega$ denotes the treewidth.
\end{theorem}

 The diameter of a directed graph is defined as the maximum length of a shortest directed path over any two vertices in any order. 
 The diameter of a directed graph that is not strongly-connected equals infinity. 
 Thus, \dmveTsc{} on directed graphs is defined as follows: given a strongly-connected directed graph $G=(V,E)$, and two integers $k$ and $\ell$, the question is whether there is a subset $F\subseteq E$ of cardinality at most $k$ such that $G-F$ is strongly-connected and $\diam(G-F)\geq \ell$?
 Observe that \dmveTsc{} on directed planar graphs parameterized by $(k,\ell)$ is FPT, as a consequence of the proof of~\cref{thm:dmvekellfpt}.

\begin{theorem}\label{thm:dirmdednopk}
 Unless~$\NP\subseteq \coNP/\poly$, \dmveTsc{} on directed planar graphs parameterized by $(k,\ell)$ does not admit a polynomial kernel. 
\end{theorem}
\begin{proof}[Proof (Sketch)]
	The following proof adapts the ideas of the proof of~\cref{thm:NoPolyKernel-D-MVE}. 
	Thus, we highlight the differences to the proof instead of providing a full proof here.
 
	We OR-cross-compose $p=2^q$ instances of the \spmveTsc{} (\spmveAcr) problem on planar, directed acyclic graphs into one instance of \dmveAcr{}$(k,\ell)$ on directed planar graphs. 
	We assume without loss of generality that in each graph of the input instances, the source reaches every vertex, and every vertex reaches the sink.
 
	We apply \cref{constr:constr2} with the following additions. 
	Let $n_{\max}$ and $k'$ be defined as in the proof of~\cref{thm:NoPolyKernel-D-MVE}, that is, $n_{\max} :=\max_{i\in[p]} |V(G_i)|$ and $k' := k^2\cdot(\log(p)+1)+k$.
	Let $L := \ell \cdot n_{\max} \cdot (2\log(p)+3)+1$ and~$\ell'=2\cdot L+\log(p)+\ell$.
	Attach to~$\lv$ as well as to~$\rv$ a path of length~$L$ each. 
	Denote the endpoint of the path attached to $\lv$ by $\lv'$ (where $\lv'\neq \lv$), and let $\rv'$ be defined analogously. 
	Direct all edges in the paths towards from~$\lv'$ to $\lv$ and from~$\rv$ to $\rv'$ respectively. 
	Moreover, add to the graph the arc $(\rv',\lv')$, and the arc $(\rv,\lv)$, the latter with cost~$k'+1$. 

	Next, we adjust the instances we compose in order to ensure that we can delete all the arcs we want without destroying the property that the source reaches every vertex and every vertex reaches the sink.
	Let $G_i$ be the graph in instance $\I_i$ for each $i\in[p]$. 
	For each arc $(v,w)\in E(G_i)$, connect $v$ and $w$ by an additional path of length $\ell$ directed towards $w$. 
	Apply this for each $G_i$, $i\in[p]$, and let $G_i'$ the graph obtained from graph $G_i$.
	Note that the directed graph $G_i'$ remains planar and acyclic. 
	Observe that none of the introduced arcs will be in a minimal solution for the \spmveAcr instance since they only occur in paths of length~$\ell$.
	Hence, $\I_i$ is a \yes-instance of \spmveAcr on planar, directed acyclic graphs if and only if $(G_i',s_i,t_i,k,\ell)$ is a \yes-instance of \spmveAcr on planar, directed acyclic graphs.
	Furthermore, in the composed \dmveTsc{}-instance, none of the introduced arcs will be deleted as this would introduce a vertex without in-going or without out-going arcs and this is not allowed in the problem setting.
	
	Let~$G$ denote the obtained graph. 
	Observe that $G$ is planar, directed and strongly-connected.

	Suppose $(G,k',\ell')$ is a \yes-instance of \dmveTsc{}.
	Consider a solution $X\subseteq E(G)$ for the instance $(G,k',\ell')$ of \dmveTsc{} on directed planar graphs. 
	The crucial observation is that for any two vertices $x,y$ not contained in the attached paths with endpoints $\lv'$ on the one, and $\rv'$ on the other hand, the following holds: $\max\{\dist_{G-X}(x,y),\dist_{G-X}(y,x)\}\leq \dist_{G-X}(\lv',\rv')$. 
	To see this, note that the arc $(\rv,\lv)$ has cost $k'+1$ and thus $(\rv,\lv)\not\in X$.  
	Since~$G$ is strongly-connected, both~$x$ and~$y$ are reachable and reach $\lv$ and $\rv$. 
	Moreover, $\lv$ is reachable from $\rv$ via the arc $(\rv,\lv)$. 
	Without loss of generality, let $\dist_{G-X}(x,y)=\max\{\dist_{G-X}(x,y),\dist_{G-X}(y,x)\}$. 
	It holds that
	\begin{align*}
		\dist_{G-X}(x,y) 	& \leq \dist_{G-X}(x,\rv) + \dist_{G-X}(\rv,\lv) + \dist_{G-X}(\lv,y) \\
								& \leq \ell \cdot n_{\max} \cdot(2\log(p)+2) + 1 + \ell \cdot n_{\max}\cdot(2\log(p)+2) \\
								& = 2 \cdot \ell \cdot n_{\max}\cdot(2\log(p)+2) + 1 < \ell'.
	\end{align*}
	Herein, recall that we allow $\log(p)+1$ arc deletions in $\vec{\triangle}_q$. 
	The second inequality follows from~\cref{lem:lengthofanyvertexdir} and the fact that in each graph~$G_i - X$ the vertex~$s_i$ has distance at most~$\ell \cdot n_{\max}$ to~$t_i$. 

	As a consequence, the vertices at distance $\ell'$ appear in the paths appended on $\lv$ and~$\rv$. 
	Among them, note that $\dist_{G-X}(\lv',\rv')$ is maximal. 
	Following the discussion in the proof of~\cref{thm:NoPolyKernel-D-MVE}, the budget has to be spend in such a way that the arc-deletions form an $\lv$-$\rv$ arc-cut in $\vec{\triangle}_q$, and the remaining budget must be spend in such a way that the instance $\I_i$ chosen by the cut allows no $s_i$-$t_i$ path of length smaller than $\ell$. 
	Hence, the $\I_i$ is a \yes-instance.

	Conversely, let $\I_i$ be a \yes-instance of \spmveAcr on planar, directed acyclic graphs and let~$X' \subseteq E(G_i)$ a minimum size solution.
	We added to each arc of $G_i$ a directed path of length $\ell$ and, as discussed above, none of the arcs in these paths is in~$X'$.
	Hence, in~$G_i - X'$ every vertex is still reachable from $s_i$ and reaches $t_i$. 
	Deleting in $G$ the arcs in~$X'$ and the arcs corresponding to the cut choosing $\I_i$ preserves the strongly-connectivity of $G$. 
	Let $X\subseteq E(G)$ be the set of deleted arcs. 
	Following the discussion in the proof of~\cref{thm:NoPolyKernel-D-MVE}, $\dist_{G-X}(\lv',\rv')\geq \ell'$.
	It follows that $\I$ is a \yes-instance of \dmveTsc{} on directed planar graphs. 
\end{proof}
\subsection{\gmve{}}\label{ss:maxgrith}

Our third fractalism application concerns the following problem.

\decprob{\gmve{} (\gmveAcr{})}
	{A directed graph $G=(V,E)$, two integers $k, \ell$.}
	{Is there a subset $F\subseteq E$ of cardinality at most $k$ such that there is no induced directed cycle of length at most $\ell$ in $G-F$?}

The problem is NP-hard~\cite{GuruswamiL14}, also on undirected graphs~\cite{Yannakakis78}.
The NP-com\-plete\-ness of \gmveAcr{} follows by a simple reduction from $k$-\textsc{Feedback Arc Set} with an $n$-vertex graph, where we set $\ell=n$ and leave the graph unchanged in the reduction. 
We remark that the problem is also known as \textsc{Cycle Transversal}~\cite{BodlaenderFLPST16}, or \textsc{$\ell$-(Directed)-Cycle Transversal}~\cite{GuruswamiL14}. 
The undirected variant is also known as \textsc{Small Cycle Transversal}~\cite{XiaZ11, XiaZ12}. 

As for the \dmveTsc problem, there is a simple search tree algorithm showing fixed-parameter tractability with respect to~$(k,\ell)$.

\begin{theorem}\label{thm:gmveisfpt}
	\gmveTsc{} can be solved in $O(\ell^k \cdot n\cdot (n+m))$ time.
\end{theorem}
\begin{proof}
We give a search tree algorithm branching over all possible edge deletions to prove that \gmveAcr{$(k,\ell)$} is fixed-parameter tractable. 
Let $(G,k,\ell)$ be an instance of \gmveAcr{}$(k,\ell)$. 
To detect short cycles in~$G$ containing a vertex~$v\in V(G)$, we construct a helping graph $G_v$ as follows. %
Delete $v$ (and all edges incident to $v$), and add $v_{\rm{in}}$ and $v_{\rm{out}}$, and the arcs $\{(x,v_{\rm{in}})\mid (x,v)\in E(G)\}$, $\{(v_{\rm{out}},x)\mid (v,x)\in E(G)\}$ as well as the arc $(v_{\rm{in}},v_{\rm{out}})$. 
Now to detect the shortest cycle in~$G$ containing $v$, compute a shortest $v_{\rm{out}}$-$v_{\rm{in}}$~path in~$G_v$. 
If a cycle is too short, then we branch into all possible, at most $\ell$ different deletions of an arc of the cycle (beside arc~$(v_{\rm{in}},v_{\rm{out}})$). 
 
 The depth of the search tree is at most $k$, and thus we obtain an $O(\ell^k\cdot n\cdot(n+m))$-time algorithm since constructing for each~$v\in V$ the helping graph~$G_v$ and then finding a shortest path in unweighted graphs can be done in~$O(n\cdot (n+m))$ time.
\end{proof}
\begin{theorem}\label{thm:noppkforgmve}
Unless~$\NP\subseteq \coNP/\poly$, \gmveTsc{} parameterized by $(k,\ell)$ does not admit a polynomial kernel.
\end{theorem}

\begin{proof}
We OR-cross-compose $p=2^q$ $\R$-equivalent instances of \spmveAcr{} on directed acyclic graphs into one instance of \gmveAcr{}($k,\ell$) as follows, where $\R$ is defined as in the proof of~\cref{theo:spmve}. 
Recall that \spmveTsc{} (\spmveAcr) on directed acyclic graphs is \NP-complete. 

\emph{Construction:} We apply~\cref{constr:constr2} with edge~cost~$k^2$. 
In addition, we add the edge~$(\rv,\lv)$ with edge~cost~$k'+1$, where~$k'=k^2\cdot(\log(p)+1)+k$. 
We denote by~$G$ the obtained graph. 
We refer to~\cref{fig:crosscomp-demg} for an exemplified illustration of the construction. 
Observe that~$G$~is not acyclic, and the edge~$(\rv,\lv)$ participates in every cycle in $G$, that is, $G$ without edge $(\rv,\lv)$ is acyclic. 
Let $(G,k',\ell')$ be the target instance of~\gmveAcr{}$(k,\ell)$ with $\ell'=\ell+\log(p)+1$. 
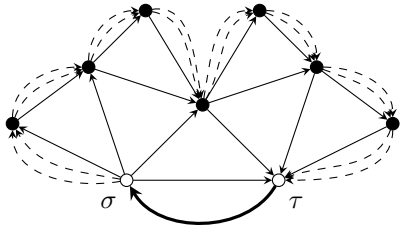
\begin{figure}[t]
\centering
 \begin{tikzpicture}

 \tikzstyle{tnode}=[fill, circle, scale=1/2, draw];
 \tikzstyle{tnodespez}=[circle, scale=1/2, draw];

\def\y{-4}

\node (d1) at (0,0+\y)[tnodespez, label=265:{$\lv$}]{};
\node (d2) at (2,0+\y)[tnodespez, label=275:{$\rv$}]{};
\node (d3) at (1,1+\y)[tnode]{};
\draw[-stealth] (d1) to (d2); 
\draw[-stealth] (d1) to (d3);
\draw[-stealth] (d3) to (d2);
\draw[-stealth, very thick] (d2) to [out=-90-40+10, in=-90+40-10](d1);

\node (d21) at (-0.5,1.5+\y)[tnode]{};
\node (d22) at (2+0.5,1.5+\y)[tnode]{};

\draw[-stealth] (d1) -- (d21);
\draw[-stealth] (d21) -- (d3);
\draw[-stealth] (d22) -- (d2);
\draw[-stealth] (d3) -- (d22);

\node (d31) at (-1.5,0.75+\y)[tnode]{};
\node (d32) at (0.25,2.25+\y)[tnode]{};
\node (d33) at (2-0.25,2.25+\y)[tnode]{};
\node (d34) at (2+1.5,0.75+\y)[tnode]{};

\draw[-stealth] (d1) -- (d31);
\draw[-stealth] (d31) -- (d21);
\draw[-stealth] (d32) -- (d3);
\draw[-stealth] (d21) -- (d32);
\draw[-stealth] (d3) -- (d33);
\draw[-stealth] (d33) -- (d22);
\draw[-stealth] (d34) -- (d2);
\draw[-stealth] (d22) -- (d34);

\draw[dashed,-stealth] (d1) to [out=170, in=-60](d31);
\draw[dashed,-stealth] (d1) to [out=190, in=-90](d31);

\draw[dashed,-stealth] (d31) to [out=70, in=-160](d21);
\draw[dashed,-stealth] (d31) to [out=90, in=-180](d21);

\draw[dashed,-stealth] (d21) to [out=70, in=-160](d32);
\draw[dashed,-stealth] (d21) to [out=90, in=-180](d32);

\draw[dashed,-stealth] (d32) to [out=-10, in=110](d3);
\draw[dashed,-stealth] (d32) to [out=10, in=100](d3);

\draw[dashed,-stealth] (d3) to [out=80, in=180](d33);
\draw[dashed,-stealth] (d3) to [out=60, in=-160](d33);

\draw[dashed,-stealth] (d33) to [out=0, in=90](d22);
\draw[dashed,-stealth] (d33) to [out=-20, in=110](d22);

\draw[dashed,-stealth] (d22) to [out=0, in=90](d34);
\draw[dashed,-stealth] (d22) to [out=-20, in=110](d34);

\draw[dashed,-stealth] (d34) to [out=-90, in=0](d2);
\draw[dashed,-stealth] (d34) to [out=-110, in=20](d2);

\end{tikzpicture}
\caption{Cross-composition for \gmveAcr{}($k,\ell$) with $p=8=2^3$. Dashed lines sketch the boundaries of the graphs in the $p$ input instances.}
\label{fig:crosscomp-demg}
\end{figure}

\emph{Correctness:} 
Note that every cycle in~$G$ uses the edge~$(\rv,\lv)$. Since its edge~cost equals $k'+1$, the budget does not allow its deletion. 
Thus, the crucial observation is that the length of any shortest path from~$\lv$ to~$\rv$ must be increased to at least~$\ell+\log(p)=\ell'-1$. 
Hence, the correctness proof follows from the proof of~\cref{thm:spmveondags}.
\end{proof}

Due to~\cref{thm:planarspmveisnphard}, using \spmveAcr{} on planar directed acyclic graphs as source problem in the proof of~\cref{thm:noppkforgmve}, we obtain the following.

\begin{theorem}\label{thm:NoPolyKernelPlanarDirectedGMVE}
	Unless~$\NP\subseteq \coNP/\poly$, \gmveTsc{} on planar directed graphs parameterized by $(k,\ell)$ does not admit a polynomial kernel.
\end{theorem}

Remarkably, \gmveAcr{}$(k,\ell)$ on planar undirected graphs admits a polynomial kernel~\cite{XiaZ11}. 
\section{Conclusion}
We start with briefly sketching how our technique can be adapt\-ed such 
that it also applies to the vertex deletion (instead of edge deletion) 
versions of the considered problems.
Afterwards, we discuss future challenges and open problems.

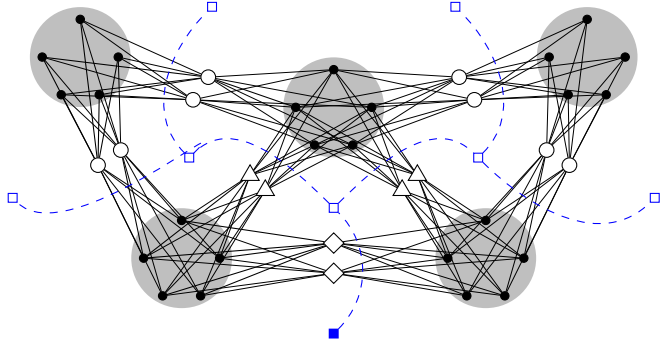
\begin{figure}
 \centering
 \begin{tikzpicture}

    \tikzstyle{tnode0}=[diamond, scale=7/12, draw];
    \tikzstyle{tnode1}=[regular polygon, regular polygon sides=3, scale=5/12, draw];
    \tikzstyle{tnode2}=[circle, scale=7/12, draw];
    \tikzstyle{tnodespez}=[circle, scale=1/2, draw];
    \tikzstyle{tnodesm}=[fill, circle, scale=1/3, draw];

    \tikzstyle{tnodet}=[rectangle, scale=1/2, color=blue, draw];
    \tikzstyle{tnodetroot}=[fill, rectangle, scale=1/2, color=blue, draw];

    \def\d{4}

    \node (d1) at (0,0)[fill, color=lightgray, circle, scale=4]{};
    \node (d2) at (\d,0)[fill, color=lightgray, circle, scale=4]{};
    \node (d3) at (\d/2,\d/2)[fill, color=lightgray, circle, scale=4]{};

    \node (d4) at (-\d/3,+\d/1.5)[fill, color=lightgray, circle, scale=4]{};
    \node (d5) at (\d+\d/3,+\d/1.5)[fill, color=lightgray, circle, scale=4]{};

    \node (d11) at (\d/2,0-0.2)[tnode0]{};
    \node (d12) at (\d/2,0+0.2)[tnode0]{};
    \node (d21) at (\d/4+0.1,\d/4-0.1)[tnode1]{};
    \node (d22) at (\d/4-0.1,\d/4+0.1)[tnode1]{};
    \node (d31) at (3*\d/4-0.1,\d/4-0.1)[tnode1]{};
    \node (d32) at (3*\d/4+0.1,\d/4+0.1)[tnode1]{};
    \node (d141) at (-\d/4-0.1,\d/3-0.1)[tnode2]{};
    \node (d142) at (-\d/4+0.2,\d/3+0.1)[tnode2]{};

    \node (d341) at (+\d/16-0.1,\d/2+0.1)[tnode2]{};
    \node (d342) at (+\d/16+0.1,\d/2+0.4)[tnode2]{};

    \node (d251) at (\d+\d/4+0.1,\d/3-0.1)[tnode2]{};
    \node (d252) at (\d+\d/4-0.2,\d/3+0.1)[tnode2]{};

    \node (d351) at (\d-\d/16+0.1,\d/2+0.1)[tnode2]{};
    \node (d352) at (\d-\d/16-0.1,\d/2+0.4)[tnode2]{};

    \node (a) at (0-0.5,0)[tnodesm]{};
    \node (b) at (0-0.25,0-0.5)[tnodesm]{};
    \node (c) at (0,0+0.5)[tnodesm]{};
    \node (d) at (0+0.25,0-0.5)[tnodesm]{};
    \node (e) at (0+0.5,0)[tnodesm]{};

    \draw[very thin] (a) to (d11);\draw (a) to (d12);
    \draw[very thin] (b) to (d11);\draw (b) to (d12);
    \draw[very thin] (c) to (d11);\draw (c) to (d12);
    \draw[very thin] (d) to (d11);\draw (d) to (d12);
    \draw[very thin] (e) to (d11);\draw (e) to (d12);

    \draw[very thin] (a) to (d21);\draw (a) to (d22);
    \draw[very thin] (b) to (d21);\draw[very thin](b) to (d22);
    \draw[very thin] (c) to (d21);\draw[very thin](c) to (d22);
    \draw[very thin] (d) to (d21);\draw[very thin](d) to (d22);
    \draw[very thin] (e) to (d21);\draw[very thin](e) to (d22);

    \draw[very thin] (a) to (d141);\draw (a) to (d142);
    \draw[very thin] (b) to (d141);\draw[very thin](b) to (d142);
    \draw[very thin] (c) to (d141);\draw[very thin](c) to (d142);
    \draw[very thin] (d) to (d141);\draw[very thin](d) to (d142);
    \draw[very thin] (e) to (d141);\draw[very thin](e) to (d142);

    \node (a) at (0-0.5+\d,0)[tnodesm]{};
    \node (b) at (0-0.25+\d,0-0.5)[tnodesm]{};
    \node (c) at (0+\d,0+0.5)[tnodesm]{};
    \node (d) at (0+0.25+\d,0-0.5)[tnodesm]{};
    \node (e) at (0+0.5+\d,0)[tnodesm]{};

    \draw[very thin](a) to (d11);\draw[very thin](a) to (d12);
    \draw[very thin](b) to (d11);\draw[very thin](b) to (d12);
    \draw[very thin](c) to (d11);\draw[very thin](c) to (d12);
    \draw[very thin](d) to (d11);\draw[very thin](d) to (d12);
    \draw[very thin](e) to (d11);\draw[very thin](e) to (d12);

    \draw[very thin](a) to (d31);\draw[very thin](a) to (d32);
    \draw[very thin](b) to (d31);\draw[very thin](b) to (d32);
    \draw[very thin](c) to (d31);\draw[very thin](c) to (d32);
    \draw[very thin](d) to (d31);\draw[very thin](d) to (d32);
    \draw[very thin](e) to (d31);\draw[very thin](e) to (d32);

    \draw[very thin](a) to (d251);\draw[very thin](a) to (d252);
    \draw[very thin](b) to (d251);\draw[very thin](b) to (d252);
    \draw[very thin](c) to (d251);\draw[very thin](c) to (d252);
    \draw[very thin](d) to (d251);\draw[very thin](d) to (d252);
    \draw[very thin](e) to (d251);\draw[very thin](e) to (d252);

    \node (a) at (0-0.5+0.5*\d,0+0.5*\d)[tnodesm]{};
    \node (b) at (0-0.25+0.5*\d,0-0.5+0.5*\d)[tnodesm]{};
    \node (c) at (0+0.5*\d,0+0.5+0.5*\d)[tnodesm]{};
    \node (d) at (0+0.25+0.5*\d,0-0.5+0.5*\d)[tnodesm]{};
    \node (e) at (0+0.5+0.5*\d,0+0.5*\d)[tnodesm]{};

    \draw[very thin](a) to (d21);\draw[very thin](a) to (d22);
    \draw[very thin](b) to (d21);\draw[very thin](b) to (d22);
    \draw[very thin](c) to (d21);\draw[very thin](c) to (d22);
    \draw[very thin](d) to (d21);\draw[very thin](d) to (d22);
    \draw[very thin](e) to (d21);\draw[very thin](e) to (d22);

    \draw[very thin](a) to (d31);\draw[very thin](a) to (d32);
    \draw[very thin](b) to (d31);\draw[very thin](b) to (d32);
    \draw[very thin](c) to (d31);\draw[very thin](c) to (d32);
    \draw[very thin](d) to (d31);\draw[very thin](d) to (d32);
    \draw[very thin](e) to (d31);\draw[very thin](e) to (d32);

    \draw[very thin] (a) to (d341);\draw (a) to (d342);
    \draw[very thin] (b) to (d341);\draw[very thin](b) to (d342);
    \draw[very thin] (c) to (d341);\draw[very thin](c) to (d342);
    \draw[very thin] (d) to (d341);\draw[very thin](d) to (d342);
    \draw[very thin] (e) to (d341);\draw[very thin](e) to (d342);

    \draw[very thin](a) to (d351);\draw[very thin](a) to (d352);
    \draw[very thin](b) to (d351);\draw[very thin](b) to (d352);
    \draw[very thin](c) to (d351);\draw[very thin](c) to (d352);
    \draw[very thin](d) to (d351);\draw[very thin](d) to (d352);
    \draw[very thin](e) to (d351);\draw[very thin](e) to (d352);

    \node (a) at (-\d/3-0.5,0+\d/1.5)[tnodesm]{};
    \node (b) at (-\d/3-0.25,0-0.5+\d/1.5)[tnodesm]{};
    \node (c) at (-\d/3,0+0.5+\d/1.5)[tnodesm]{};
    \node (d) at (-\d/3+0.25,0-0.5+\d/1.5)[tnodesm]{};
    \node (e) at (-\d/3+0.5,0+\d/1.5)[tnodesm]{};

    \draw[very thin] (a) to (d142);\draw (a) to (d141);
    \draw[very thin] (b) to (d142);\draw (b) to (d141);
    \draw[very thin] (c) to (d142);\draw (c) to (d141);
    \draw[very thin] (d) to (d142);\draw (d) to (d141);
    \draw[very thin] (e) to (d142);\draw (e) to (d141);

    \draw[very thin] (a) to (d341);\draw (a) to (d342);
    \draw[very thin] (b) to (d341);\draw[very thin](b) to (d342);
    \draw[very thin] (c) to (d341);\draw[very thin](c) to (d342);
    \draw[very thin] (d) to (d341);\draw[very thin](d) to (d342);
    \draw[very thin] (e) to (d341);\draw[very thin](e) to (d342);

    \node (a) at (\d+\d/3-0.5,0+\d/1.5)[tnodesm]{};
    \node (b) at (\d+\d/3-0.25,0-0.5+\d/1.5)[tnodesm]{};
    \node (c) at (\d+\d/3,0+0.5+\d/1.5)[tnodesm]{};
    \node (d) at (\d+\d/3+0.25,0-0.5+\d/1.5)[tnodesm]{};
    \node (e) at (\d+\d/3+0.5,0+\d/1.5)[tnodesm]{};

    \draw[very thin](a) to (d351);\draw[very thin](a) to (d352);
    \draw[very thin](b) to (d351);\draw[very thin](b) to (d352);
    \draw[very thin](c) to (d351);\draw[very thin](c) to (d352);
    \draw[very thin](d) to (d351);\draw[very thin](d) to (d352);
    \draw[very thin](e) to (d351);\draw[very thin](e) to (d352);

    \draw[very thin](a) to (d251);\draw[very thin](a) to (d252);
    \draw[very thin](b) to (d251);\draw[very thin](b) to (d252);
    \draw[very thin](c) to (d251);\draw[very thin](c) to (d252);
    \draw[very thin](d) to (d251);\draw[very thin](d) to (d252);
    \draw[very thin](e) to (d251);\draw[very thin](e) to (d252);

    \node (t1) at (\d/2,-\d/4)[tnodetroot]{};
    \node (t2) at (\d/2,+\d/6)[tnodet]{};

    \node (t3) at (0.1,+\d/3)[tnodet]{};
    \node (t4) at (\d-0.1,+\d/3)[tnodet]{};

    \node (t5) at (0.4,+\d/1.2)[tnodet]{};
    \node (t6) at (\d-0.4,+\d/1.2)[tnodet]{};

    \node (t7) at (-\d/1.8,+\d/5)[tnodet]{};
    \node (t8) at (\d+\d/1.8,+\d/5)[tnodet]{};

    \draw[dashed, color=blue] (t1) to [out=45,in=-45](t2);
    \draw[dashed, color=blue] (t2) to [out=135,in=45](t3);
    \draw[dashed, color=blue] (t2) to [out=45,in=135](t4);
    \draw[dashed, color=blue] (t3) to [out=135,in=-135](t5);
    \draw[dashed, color=blue] (t3) to [out=45,in=-45](t7);
    \draw[dashed, color=blue] (t4) to [out=45,in=-45](t6);
    \draw[dashed, color=blue] (t4) to [out=-45,in=-135](t8);

    \end{tikzpicture}

\caption{The vertex deletion variant~$\triangle_2^{2;5}$ of \fractal{}s. Vertex types: empty diamonds belong to the boundary~$B_0$, empty triangles belong to the boundary~$B_1$, empty circles belong to the boundary~$B_2$. The squares and dashed lines indicate the dual structure, where the filled square corresponds to the root. We highlighted vertices in gray-filled circles that correspond to the vertices in the edge-deletion variant~$\triangle_2$. }
\label{fig:edgedeltovertexdel}
\end{figure}

\paragraph*{Vertex-Deletion Variants.}%

We give another modification of the \fractal{} such that vertex-deletion variants can be tackled. 
We obtain the vertex-deletion variant $\triangle_q^{c;d}$ of the~\fractal{} from $\triangle_q^{c}$ as follows, where $d$ denotes an additional \emph{vertex~cost}. 
Recall that $\triangle_q^{c}$ can be reduced to an unweighted, simple graph~$\hat{\triangle}_q^{c}$. 
We first obtain $\hat{\triangle}_q^{c}=(V'\cup V'', E')$ from $\triangle_q^{c}$, where $V'$ denote the vertices not being the product of a subdivision in the step from $\triangle_q^{c}$ to $\hat{\triangle}_q^{c}$. 
Next, we describe how to obtain $\triangle_q^{c;d}$ from $\hat{\triangle}_q^{c}$. 
To this end, we introduce the following notation: given a graph~$G=(V,E)$ and $v\in V$, we say we \emph{clone} vertex $v$ if we add a new vertex $v'$ to $V$ and the edge~set $\{\{v',w\}\mid\{v,w\}\in E\}$ to $E$. 
We obtain $\triangle_q^{c;d}$ from $\hat{\triangle}_q^{c}$ by cloning every vertex in $V'$ $d-1$~times (we refer to them as the \emph{clones} in the following). 
We denote by $C_x\subseteq V(\triangle_q^{c;d})$ the clones of vertex $x\in V(\hat{\triangle}_q^{c})$. 
We refer to \cref{fig:edgedeltovertexdel} for an illustration of the vertex-deletion variant of \fractal{} $\triangle_2$ with edge~cost 2 and vertex~cost 5.

The vertex~cost can be interpreted as a tool to avoid deletion of clones. 
Herein, we can set the vertex~cost larger than the budget for vertex-deletions in a given problem instance to avoid any deletion of clones. 
To this end, note that to essentially change the structure of the graph by deleting a vertex having clones, it is required the delete all clones of the vertex as well.

We remark that the vertex-deletion variant of the~\fractal{} can be directed in the same way as the edge-deletion variant of the~\fractal{} such that the obtained graph is acyclic. 
Moreover, we can transfer the notion of boundaries, now being a set of vertices instead, as well as the dual structure for the vertex-deletion variant of~\fractal{} (cf.~\cref{fig:edgedeltovertexdel}). 
Note that in general~$\triangle_q^{c;d}$ is not planar, for example for~$c,d\geq 3$.

One can show that all properties of the edge-deletion variant also hold on the vertex-deletion variant, replacing edge~cuts by vertex~cuts (modulo some constants), while forbidding to delete clones.
Again, the latter is reasonable since in any application we can set the vertex~cost larger than the budget for vertex-deletions.
For example, considering any minimum $C_\lv$-$C_\rv$~vertex cut in~$\triangle_q^{c;d}$, where every vertex in every~$C_x$, $x\in V(\triangle_q)$, is not allowed to be deleted. 
One can show that it is of size $(q+1)\cdot c$, using a simple bijection of the edges in~$\triangle_q^c$ and the corresponding vertices in~$\triangle_q^{c;d}$. 

In addition, one can modify~\cref{constr:constr1,constr:constr2} slightly to use the vertex-deletion variants for vertex-deletion problems. 
Herein, it is worth to mention how the merging of the source and sink vertices of the input instances works. 
Consider $s_i$ and $v_{i-1}$ as defined in~\cref{constr:constr1}, and let $d\in\N$ be the vertex~cost. 
Note that $v_{i-1}$ is replaced by $C:=C_{v_{i-1}}$ with $|C|=d$. 
We remove $s_i$ and all incident edges of $s_i$, and add $d$ copies $s_i^{1},\ldots,s_i^{d}$ of $s_i$. 
In addition, if $\{s_i,x\}$ was an edge we deleted in the previous step, we add the edges $\{s_i^{j},x\}$ for all $j\in[d]$. 
Finally, we merge each $s_i^{j}$ with one vertex in~$C$ in such a way that each vertex in~$C$ is merged exactly once. 
We apply an analogue procedure to the sink vertex~$t_i$ and~$C_{v_i}$. 

We remark that recently, Zschoche~\cite{Zschoche17} used the fractalism technique to exclude the existence of polynomial kernels for the problem of finding $s$-$t$ separators in temporal graphs under $\NP\not\subseteq \coNP/\poly$.
Moreover, a gadget relying on grids that could ensure planarity for the vertex-deletion variant of the \fractal{} is proposed.

\subparagraph*{Outlook.}
We provided several case studies where our fractalism technique applies.
It remains open to further explore the limitations and possibilities of our technique in more contexts.
Note that the fractal structure also applies for refuting the existence of polynomial kernels for problems not dealing with cuts~\cite{GuillemotHPP13}.
\cref{tab:overview} in \cref{sec:intro} presents an open question which should be clarified. %
Moreover, we could not settle the cases for vertex deletion problems when the 
underlying graphs are planar.

\section*{Acknowledgement}
We thank Manuel Sorge (TU Berlin) for his help with the proof of~\cref{thm:planarspmveisnphard}.

\bibliographystyle{plain}
\bibliography{fractal-arxiv}

\end{document}